\documentclass[a4paper,11pt,aps, prd]{revtex4-1}
\pdfoutput=1
\usepackage{amsmath}
\usepackage{amssymb}
\usepackage{float}
\usepackage[T1]{fontenc} 
\usepackage{tikz}
\newcommand{\redlinedashed}{\raisebox{2pt}{\tikz{\draw[-,red,densely dashed,line width = 0.2pt](0,0) -- (8mm,0);}}}
\newcommand{\redline}{\raisebox{2pt}{\tikz{\draw[-,red,line width = 0.2pt](0,0) -- (8mm,0);}}}
\newcommand{\orangeline}{\raisebox{2pt}{\tikz{\draw[-,orange,line width = 0.2pt](0,0) -- (8mm,0);}}}
\begin{document}
\title{\boldmath Thermodynamic instabilities in holographic neutron stars at finite temperature}
\author{Carlos R. Arg\"uelles$^{a,b}$}\email{carguelles@fcaglp.unlp.edu.ar}
\author{Tob\'\i as Canavesi$^{a,c}$}\email{tcanavesi@fisica.unlp.edu.ar}
\author{Manuel D\'\i az$^c$}\email{mnig.diaz@gmail.com}
\author{and Nicol\'as Grandi$^{c,b}$}\email{grandi@fisica.unlp.edu.ar}
\affiliation{$^a$Facultad de Ciencias Astron\'omicas y Geof\'isicas de La Plata,\\Paseo del Bosque s/n, B1900FWA La Plata, Argentina}
\vspace{.5cm}
\affiliation{$^b$Departamento de F\'\i sica, Universidad Nacional de La Plata,\\Calle 49 y 115 s/n, CC67, 1900 La Plata, Argentina}
\vspace{.5cm}
\affiliation{$^c$Instituto de F\'isica de La Plata, \\
\mbox{Diagonal 113 63 y 64, CC67, 1900 La Plata, Argentina}
}
\begin{abstract}
\noindent {\bf Abstract:} 
We study the thermodynamics of a self-gravitating system of neutral fermions at finite temperature and analyze its backreaction in an asymptotically AdS space. We evaluate numerically the free entropy as a function of temperature, and perform a stability analysis applying the simpler and powerful graphical method referred as the Katz criterion. We found that for highly-enough degenerate fermionic solutions, the onset of thermodynamic instability arises, though prior to the turning point on the mass as a function of the central density. Our results for finite temperature fermions provide a novel and more general way to study the confinement to deconfinement phase transition in the holographic field theory, generalizing former conclusions developed for systems at zero temperature.
\end{abstract} 

\maketitle

\section{Introduction}
\label{sec:introduction}
The thermodynamics of self-gravitating systems is a fascinating subject with more than 50 years of related research \cite{1990PhR...188..285P, Katz2003, 2006IJMPB..20.3113C}. The particularity introduced by the long range and unshielded nature of gravitational force is that equilibrium states become non-hom\-ogeneous, and the resulting thermodynamics is not extensive \cite{2006IJMPB..20.3113C}. Indeed, due to the nonadditivity of the energy in the presence of long range interactions, there is an inequivalence of ensembles in self-gravitating systems \cite{1990PhR...188..285P, 2009PhR...480...57C}, contrasting with the more traditional thermodynamics of gases or plasmas. In particular, the thermodynamic properties of a self-gravitating neutral fermionic fluid, or ``neutron star'', were first described in \cite{1970ZPhy..235..339T} within Newtonian gravity, and many years later within general relativity in \cite{1999EPJC...11..173B}. The availability of more powerful numerical resources in recent years, allowed for a more detailed description of the different phases and their thermodynamic stability in different ensembles \cite{2006IJMPB..20.3113C,2015PhRvD..92l3527C,2019arXiv190810303C}. Regarding the dynamical equilibrium, the Tolman-Oppenheimer-Volkoff equations \cite{T,OV,1930PhRv...35..904T}  for a perfect fermionic fluid at finite temperature, present a rich set of solutions as a function of the central temperature and the central degeneracy \cite{1990A&A...235....1G}. The dynamical stability of such systems was first studied in  \cite{1984MNRAS.207P..13C}.  Such states develop a ``massive core - diluted halo'' structure as the central degeneracy is increased at fixed central temperature (see \cite{2006IJMPB..20.3113C, 2015PhRvD..92l3527C} for the case of Newtonian gravity, and \cite{2019arXiv190810303C, 2018arXiv180801007A} for the general relativistic setup), until the central core reaches the critical point of gravitational collapse \cite{2014IJMPD..2342020A}.

These general relativistic studies have phenomenological as well as theoretical applications. From the phenomenological perspective, in \cite{2019arXiv190810303C}  such fermionic mass distributions were used to model massive stars such as red giants or supernova with a possible subsequent collapse to a black hole in the latter. On the other hand in \cite{2015MNRAS451622R,  2015ARep...59..656S,2016PhRvD..94l3004G, 2016JCAP...04..038A,2018PDU....21...82A, 2019PDU....24..278A, 2019arXiv190509776A} the ``core - halo'' configurations where used to model dark matter halos in galaxies, providing rotation curves that are well fitted to observations, while the central core is able to mimic the massive black hole at the galactic centers for KeV particle masses. From the theoretical side on the other hand, in \cite{2018JHEP...05..118A} an AdS asymptotics was imposed, in order to interpret the system as a ``holographic neutron star'' \cite{DeBoer2009, Arsiwalla2010}, and scalar correlators of the dual field theory were calculated.

In this work, we explore the AdS construction, with emphasis on the thermodynamic stability of the fermion gas, using the rigorous stability criterion first developed by Katz \cite{1978MNRAS.183..765K} for classical self-gravitating systems in flat space. The main results presented on \cite{2018JHEP...05..118A} were that $(i)$ the core-halo profiles also show up in the AdS case, and $(ii)$ the scalar correlators of the boundary theory develop a ``swallow tail'' structure as the central degeneracy in increased. Here, we explore further this system, with particular interest in its thermodynamic instabilities. 

Our main goal is to calculate the grand canonical potential and free entropy of the AdS solutions and to express them as a function of the boundary temperature and chemical potential - instead of the central quantities. Our main result is to identify characteristic features of the unstable regions that can be used as proxies to diagnose instability in this kind of systems. We show that the turning point on the mass as a function of the central density occurs inside the unstable regions. 

We concentrate in the holographic approach, in which the gravitational solution is not intended to represent a real astrophysical object, but is used instead as a calculational device to obtain information on the strongly coupled conformal field theory defined on the holographic boundary. In this context, the neutron star solution is dual to a highly degenerate fermionic state of the boundary field theory \cite{DeBoer2009, Arsiwalla2010}. Since it asymptotes global AdS, the boundary theory lives on a finite volume sphere. This is in contrast with the planar configurations studied in \cite{Hartnoll:2010gu, Hartnoll:2010ik, Puletti:2010de}, in which the boundary theory is defined in a two-dimensional flat Euclidean space. In the holographic context, the gravitational collapse of the bulk solution is interpreted as the confinement to deconfinement phase transition of the boundary fermions. Thus, our results provide a novel and more general way to study such phase transition of the boundary theory, while generalizing former conclusions developed for systems at zero temperature and/or infinite volume.

\newpage

\section{The bulk perspective}
\label{sec:bulk1}
We work on the neutron star background in asymptotically globally AdS spacetime, originally constructed in \cite{DeBoer2009, Arsiwalla2010} and extended to finite temperature in \cite{2018JHEP...05..118A}. 

\subsection{Building the bulk state: a self-gravitating fluid at finite temperature}
\label{sec:background1}
In this section, we sketch the steps of the construction of the corresponding bulk state, originally published in the aforementioned references. For the reader's convenience, we include the detailed derivation in Appendix \ref{sec:background}.

We want to solve Einstein equations with a stationary spherically symmetric and asymptotically AdS ``neutron star'' Ansatz 
\begin{eqnarray}\label{metric1}
&&ds^{2}=L^{2}\left(-e^{\chi } \left( 1-\frac{2\tilde{M}}{r} +r^2 \right)dt^{2}+\frac{dr^{2}}{ 1-\frac{2\tilde{M}}{r} +r^2 }+r^2d\Omega^2_2\right),
\end{eqnarray}
where $L$ is the AdS length, $d\Omega^2_2=d\vartheta^2+\sin^2\!\vartheta\, d\varphi^2$ is a two-sphere, and the dimensionless mass $\tilde{M}(r)$ and the exponent $\chi(r)$ in the lapse function  depend on the radius $r$. The resulting Einstein equations read
\begin{equation}\label{eq:M1}
\frac{d\tilde{M}}{d r}=4\pi r^2\tilde{\rho},
\qquad\qquad\qquad
\frac{d {\chi}}{dr}=\frac{8\pi r\left(\tilde{P}+\tilde{\rho}\right)}{ 1-\frac{2\tilde{M}}{r} +r^2 }\,,
\end{equation}
in terms of the fluid dimensionless density $\tilde \rho$ and pressure $\tilde P$, that are obtained from  the expressions corresponding to a very large number of neutral fermions in local thermodynamic equilibrium, as
\begin{equation}
\label{eq:density.dimensionless1}
\tilde \rho=\gamma^2\int_1^\infty
\frac{\epsilon^2\sqrt{\epsilon^2-1}}{e^{\frac{\epsilon-\tilde{\mu}}{\tilde{T}}}+1}\,d\epsilon\,,
\qquad\qquad\qquad
\tilde P=
\frac{\gamma^2}3\int_1^\infty
\frac{\left(\sqrt{\epsilon^2-1}\right)^3}{e^{\frac{\epsilon-\tilde{\mu}}{\tilde{T}}}+1}\,d\epsilon\,,
\end{equation}
where $\gamma^2$ is a coupling constant. Here the local dimensionless temperature $\tilde T$ and chemical potential $\tilde\mu$ are radial functions that satisfy the thermodynamic equilibrium conditions of Tolman and Klein
\begin{equation}\label{eq:tolman1}
e^{\frac{\chi}2 } \left( 1-\frac{2\tilde{M}}{r} +r^2 \right)^{\frac{1}{2}}\tilde{T}={\rm constant}\,,
\qquad\qquad\qquad
e^{\frac{\chi}2 } \left( 1-\frac{2\tilde{M}}{r} +r^2 \right)^{\frac{1}{2}}\tilde{\mu}={\rm constant}\,.
\end{equation}

By expanding the equations \eqref{eq:M1}  around the center of the configuration $r=0$, we obtain the boundary conditions that correspond to a regular metric, in the form
\begin{eqnarray}
&&\tilde{M}(0)=0,
\qquad\qquad\qquad\qquad
\chi(0)=0,\nonumber\\
&&\tilde{T}(0)=\tilde{T_{0}},
\qquad\qquad\qquad\qquad
\tilde{\mu}(0)=\tilde \mu_{0}\equiv \Theta_{0}\tilde{T_{0}}+1.
\label{eq:boundary.mu1}
\end{eqnarray}
Here $\Theta_{0}$ is called the ``central degeneracy'', and we used it as a way to parameterize the central chemical potential. Families of solutions are then indexed by the parameters $(\tilde T_0, \Theta_0,  \gamma^2)$.
 
\subsection{Probing the bulk state: scattering of an Euclidean massive particle}
\label{subsec:geodesics1}
 
In order to probe the resulting gravitational background, we study massive space-like geodesics starting and ending at the AdS boundary. 
The details of the derivation are given in Appendix \ref{subsec:canonicalpotential1}, since they were previously discussed in \cite{2018JHEP...05..118A}.
 

We focus in geodesics starting at a very large radius $r_\epsilon$, falling into the geometry up to a minimum radius $r_*$, and then bouncing back into the asymptotic region. The time $t$ is kept fixed, and the initial and final points span a total angle $\Delta \varphi$ at the boundary. 
%
This is a function of $r_*$ which reads
\begin{equation}
\Delta\varphi = 2r_*\int_{r_*}^{r_\epsilon} dr\,\frac{ e^{\frac{\lambda(r)}2}}{
r \sqrt{
r^2-r_*^2
}
}\,.
\label{eq:delta.varphi1}
\end{equation}
For $r_*=0$ we have no scattering and then $\Delta\varphi=\pi$. On the other hand in the limit of very large  $r_*$ we get $\Delta\varphi=0$, implying a backward scattering. In the intermediate region, the behavior of $\Delta\varphi$ can be either monotonic or non-monotonic. In the last case, the same angle $\Delta\varphi$ is spanned by geodesics with different values of the minimum approach radius $r_*$.

\begin{figure}[ht]
\centering
\vspace{-1.7cm}
\includegraphics[width=0.6\textwidth]{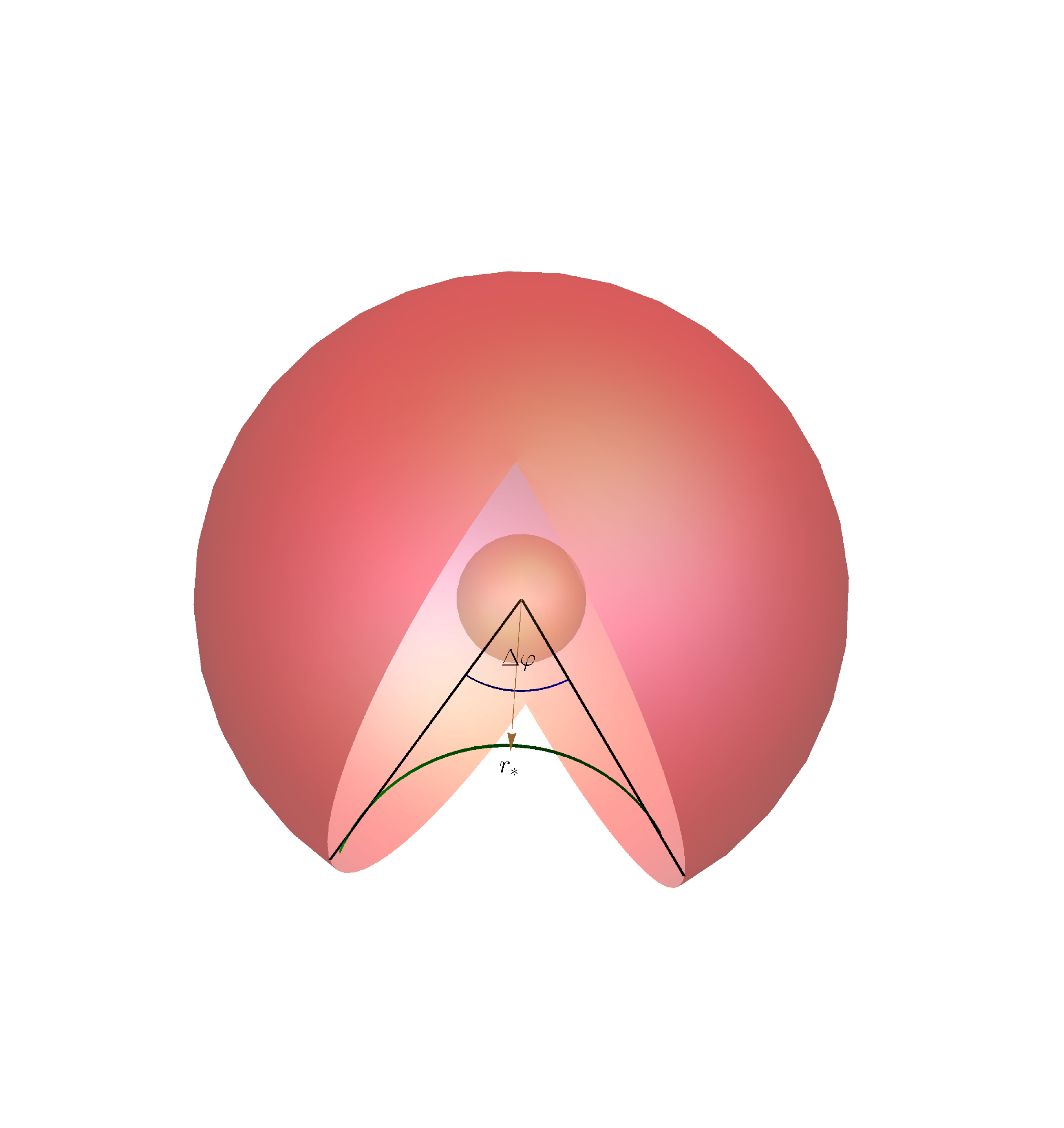}
\vspace{-1.7cm}
\caption{\label{fig:sphere} We study the scattering problem of a massive Euclidean particle entering the geometry from infinity, approaching the neutron star up to a tip radius $r_*$, and then moving again to the asymptotic region, spanning an angle $\Delta \varphi$.}
\end{figure}

\newpage 

\subsection{Results from the bulk perspective}
\begin{figure}[H]
\centering
\includegraphics[width=0.45\textwidth]{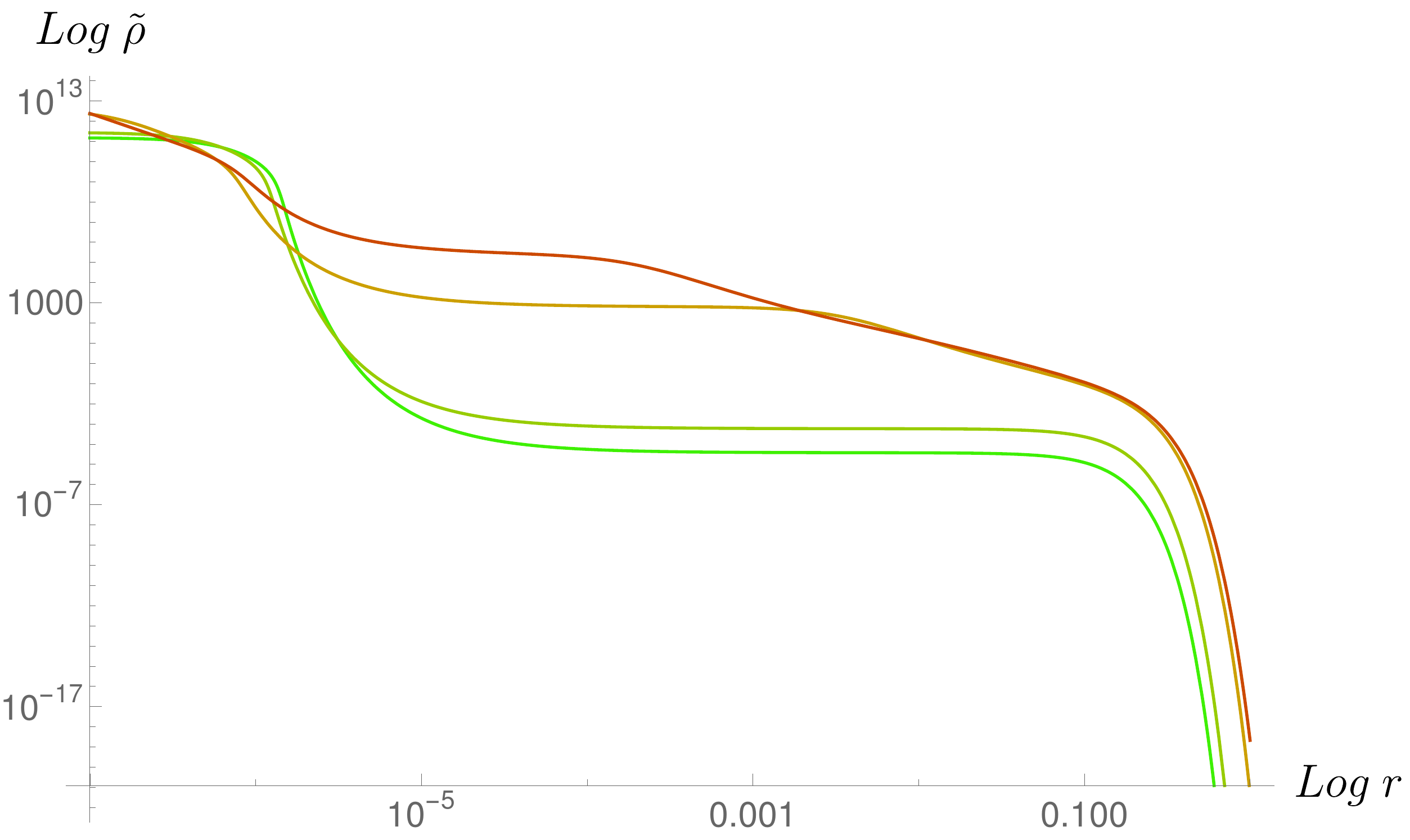}
\hfill
\includegraphics[width=0.45\textwidth]{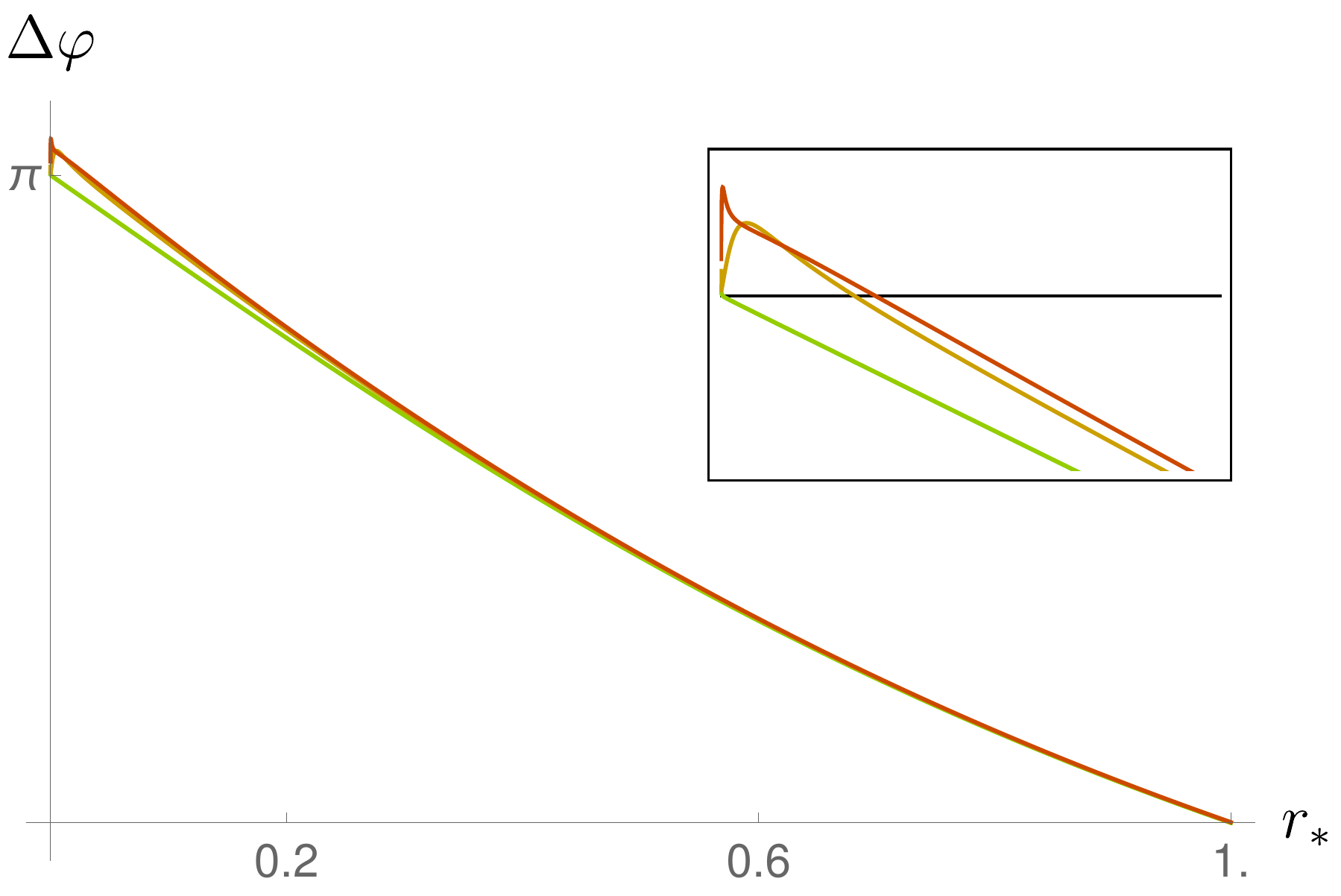}
\hfill
\\
\includegraphics[width=0.45\textwidth]{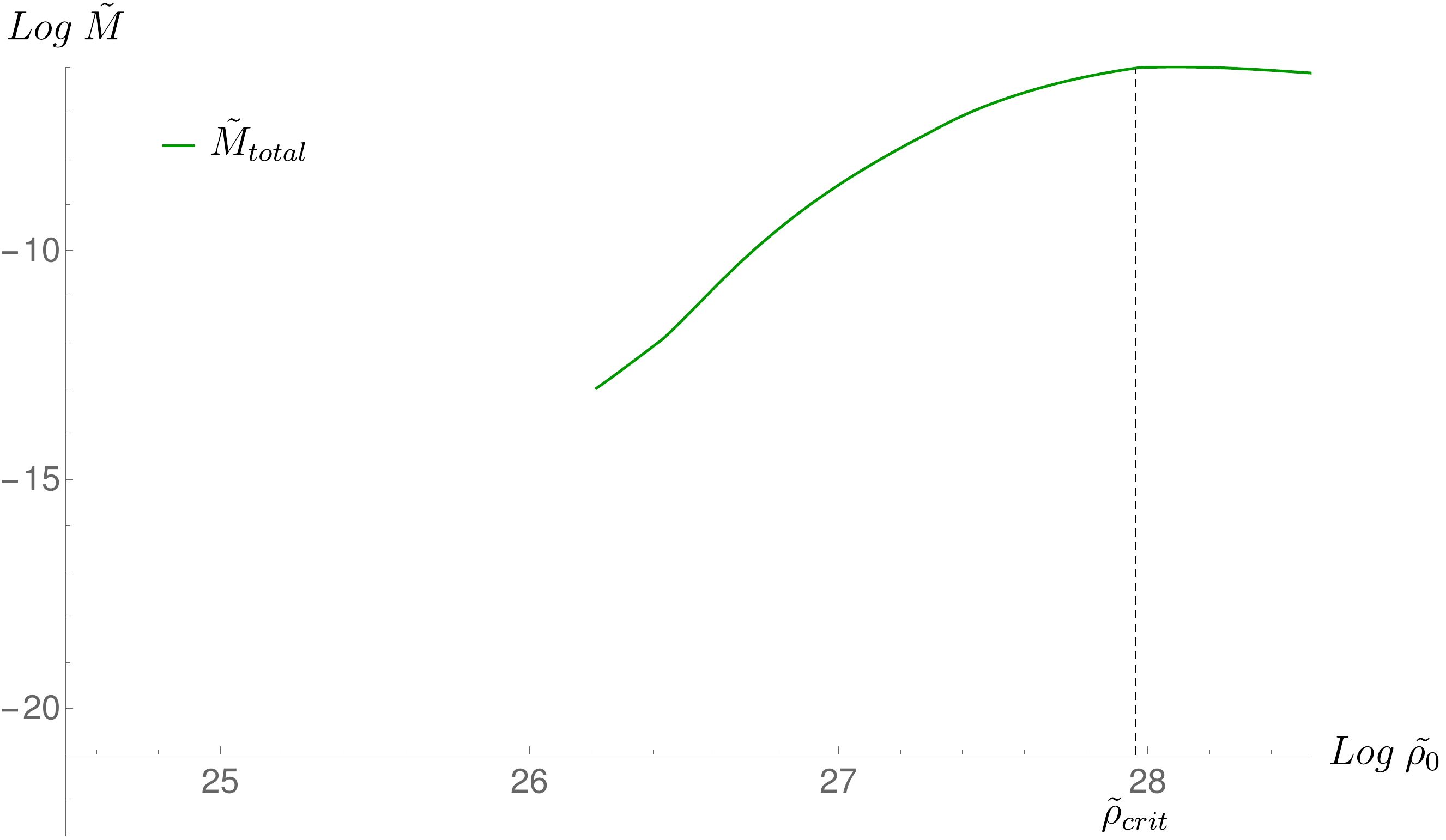}
\hfill
\includegraphics[trim=-5mm 0mm 0 10mm,width=0.2\textwidth]{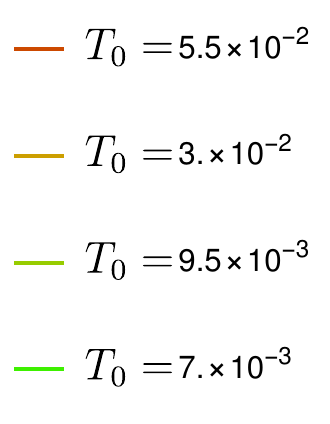}\hfill
\put(-180,30){\scriptsize$\Theta_0=50$\normalsize}
\put(-180,10){\scriptsize$\gamma=7\times10^5$\normalsize}
\put(-60,90){$\tilde{\phantom T}$}
\put(-60,64){$\tilde{\phantom T}$}
\put(-60,37){$\tilde{\phantom T}$}
\put(-60,11){$\tilde{\phantom T}$}
\\
\centerline{-------- o --------}
\includegraphics[width=0.45\textwidth]{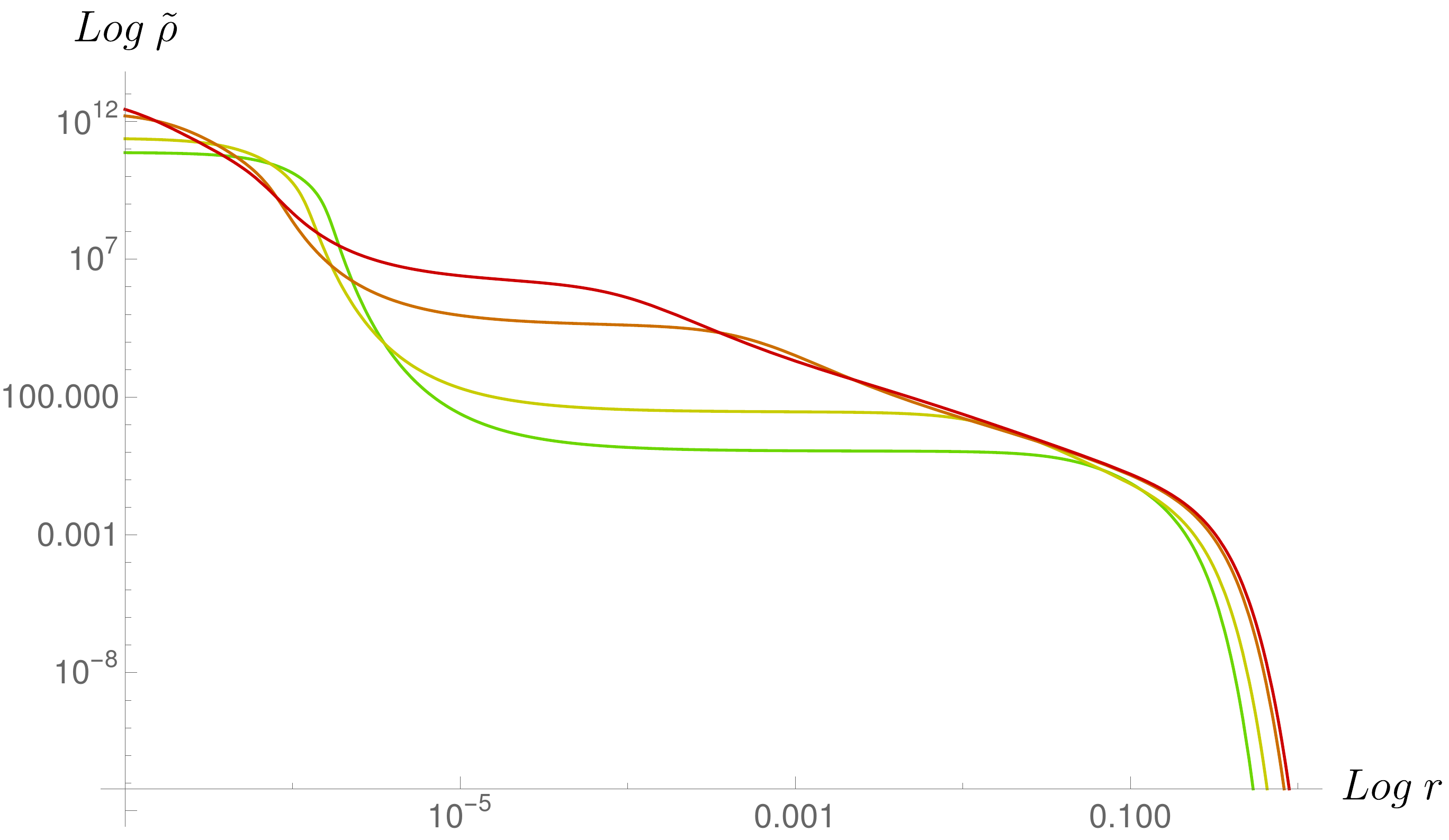}
\hfill
\includegraphics[width=0.45\textwidth]{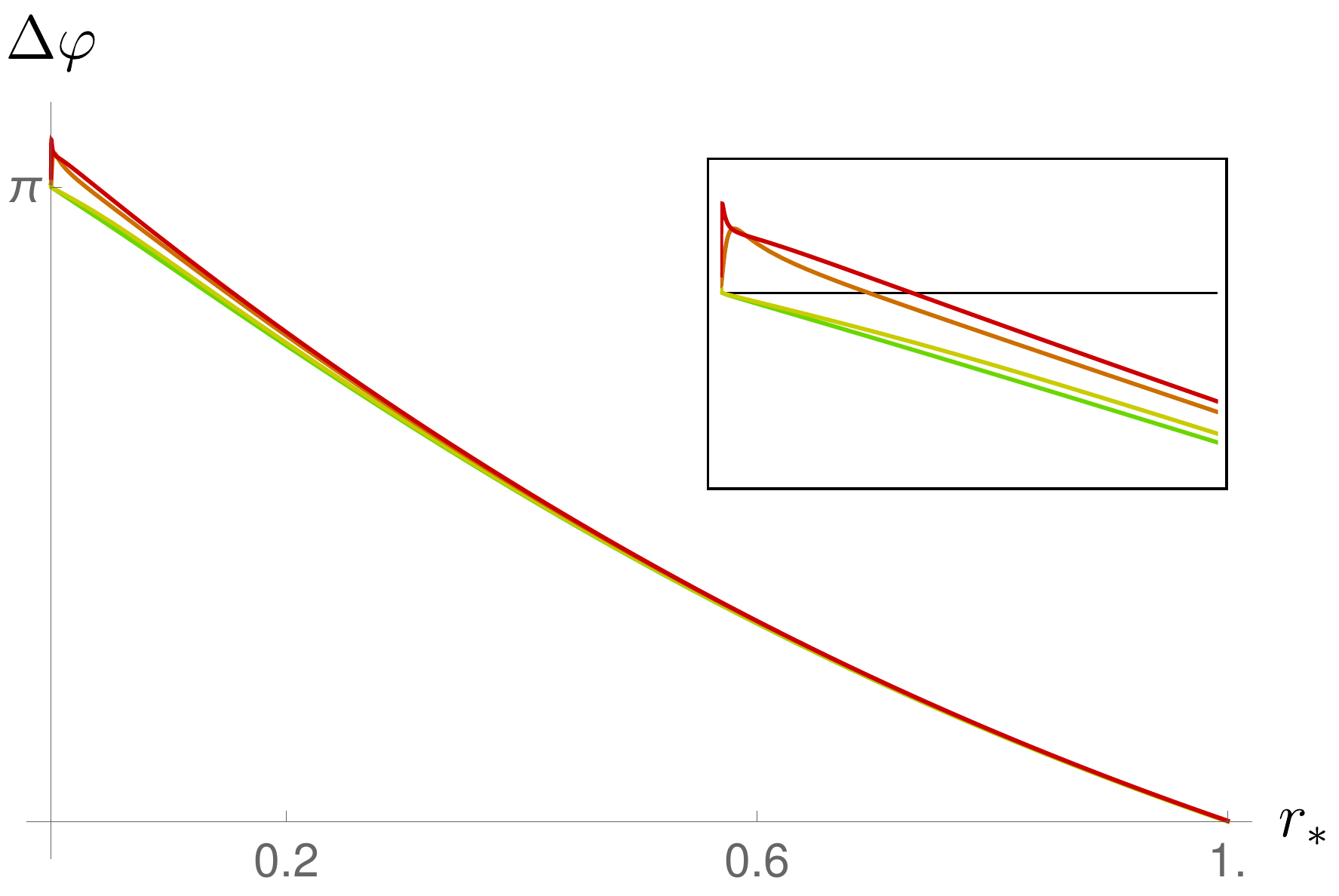}
\hfill
\\
\includegraphics[width=0.45\textwidth]{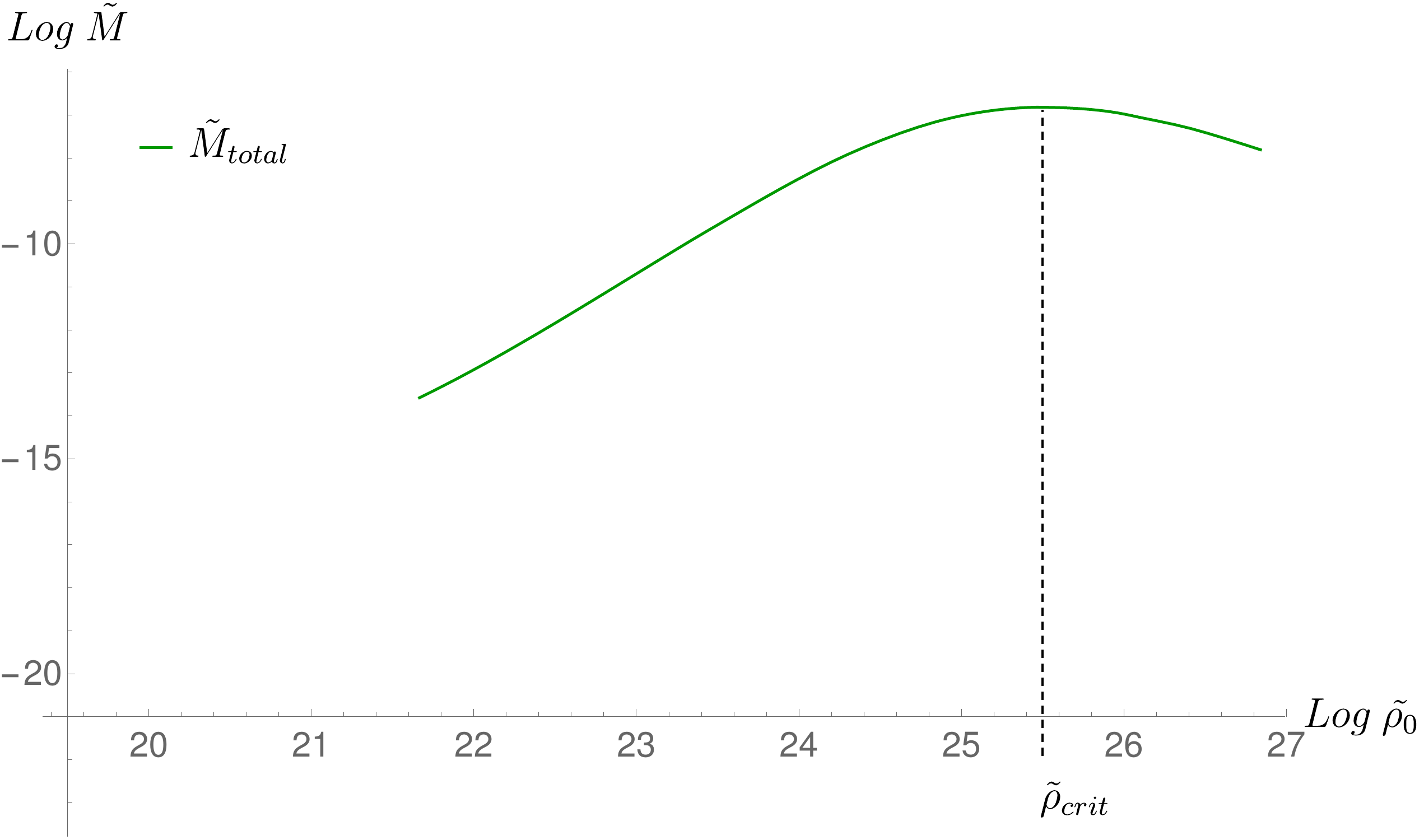}
\hfill
\includegraphics[trim=-5mm 0mm 0 10mm,width=0.2\textwidth]{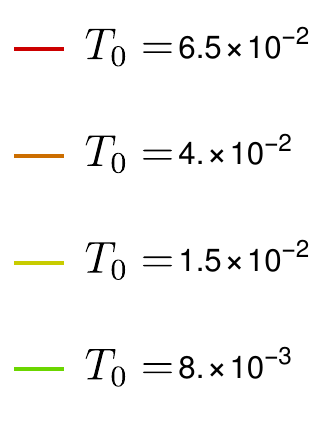}	
\put(-180,30){\scriptsize$\Theta_0=30$\normalsize}
\put(-180,10){\scriptsize$\gamma=7\times10^5$\normalsize}
\put(-60,90){$\tilde{\phantom T}$}
\put(-60,64){$\tilde{\phantom T}$}
\put(-60,37){$\tilde{\phantom T}$}
\put(-60,11){$\tilde{\phantom T}$}
\caption{\label{perfil50} Plots of the solutions corresponding to $\Theta_0=50$ (first set of three plots) and $\Theta_0=30$ (second set of three plots) for different values of $\tilde {T}_0$.
\underline{Up-left:} logarithmic plot of the density profile as a function of the radius. The density has a dense core  and a diluted halo that decreases sharply at the boundary of the star for low temperatures, while for $\tilde{T}_0> \tilde{T}_0^{cr}$  it takes a power law form. \underline{Up-right:} the angle $\Delta\varphi$ between the incident and scattered direction of a massive Euclidean geodesic. For the profiles with a power law decay, it grows into values bigger than $\pi$, as can be seen in the inset.
\underline{Down-left:} the total mass $\tilde M$ as a function of the central density. We see that there is a local maximum, approximately at the same temperatures at which the power law behavior of the star edge and the non-monotonic $\Delta \varphi$ show up.}
\label{fig:positivetheta1}
\end{figure}

\begin{figure}[H]
\centering
\includegraphics[width=0.45\textwidth]{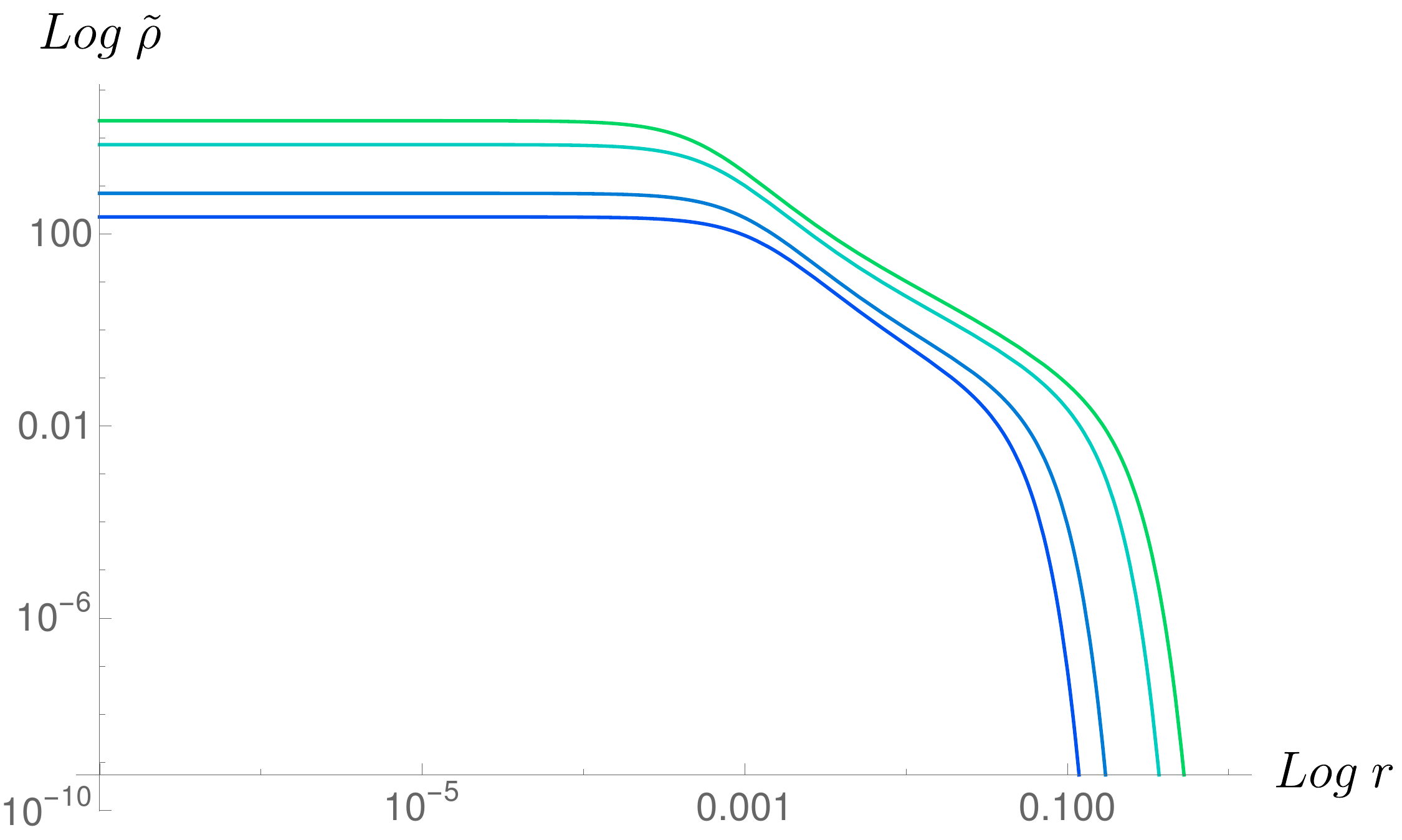}
\hfill
\includegraphics[width=0.45\textwidth]{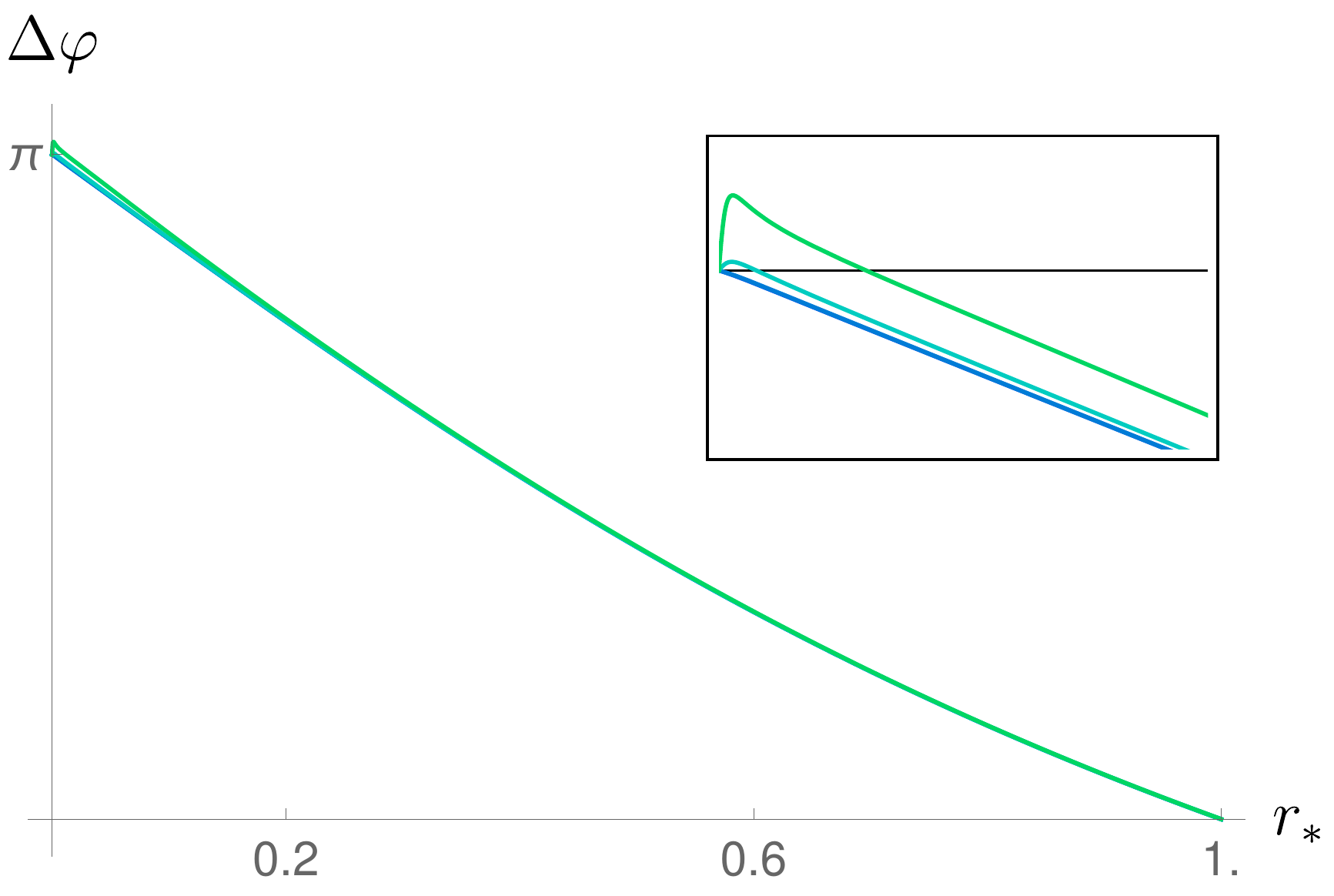}
\hfill
\\
\includegraphics[width=0.45\textwidth]{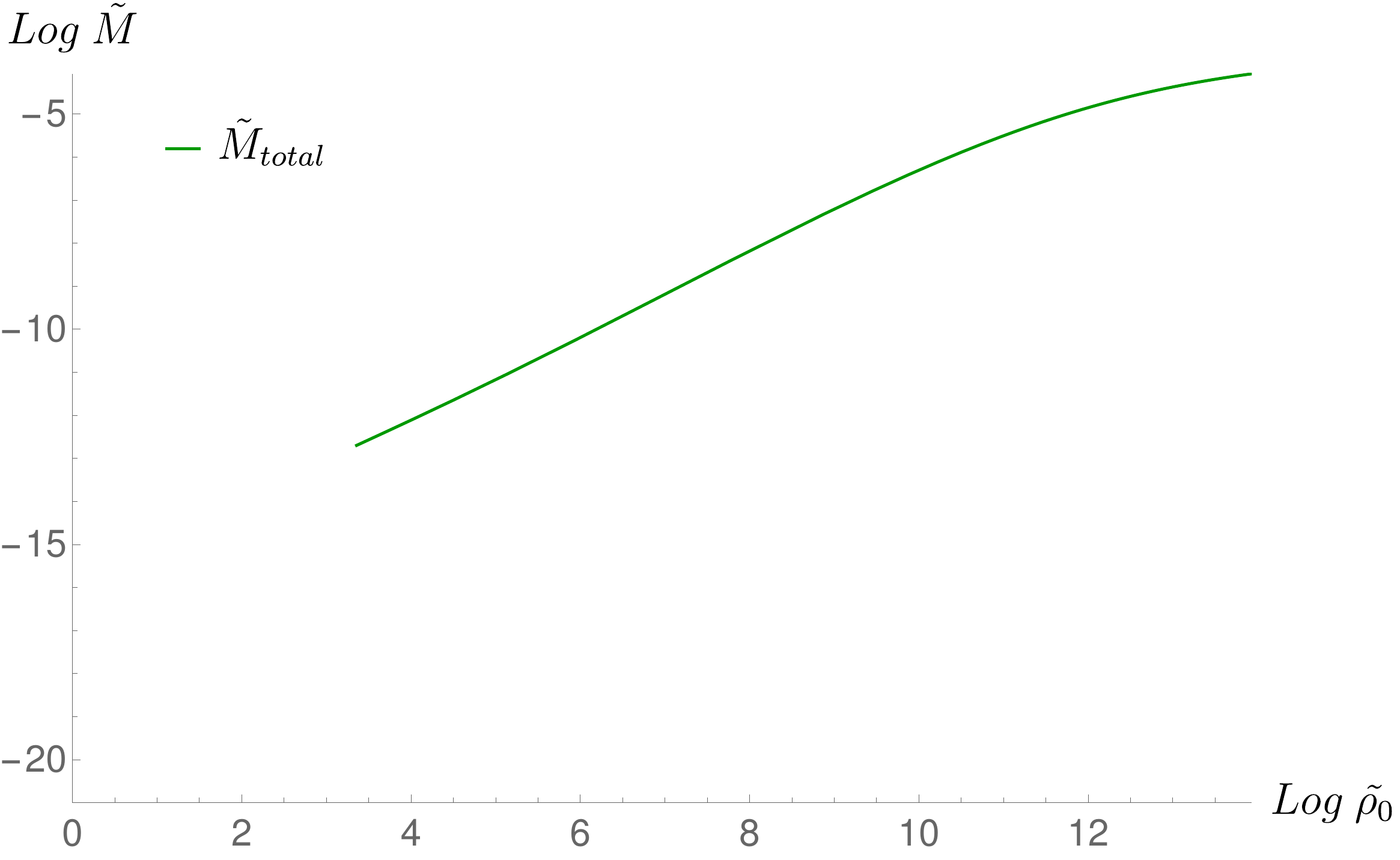}
\hfill
\includegraphics[trim=-5mm 0mm 0 10mm,width=0.2\textwidth]{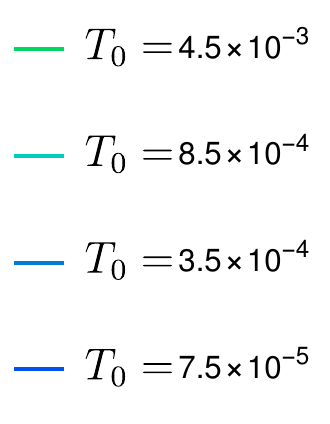}	
\put(-180,30){\scriptsize$\Theta_0=-10$\normalsize}
\put(-180,10){\scriptsize$\gamma=7\times10^5$\normalsize}
\put(-60,90){$\tilde{\phantom T}$}
\put(-60,64){$\tilde{\phantom T}$}
\put(-60,37){$\tilde{\phantom T}$}
\put(-60,11){$\tilde{\phantom T}$}
\\
\centerline{-------- o --------}
\includegraphics[width=0.45\textwidth]{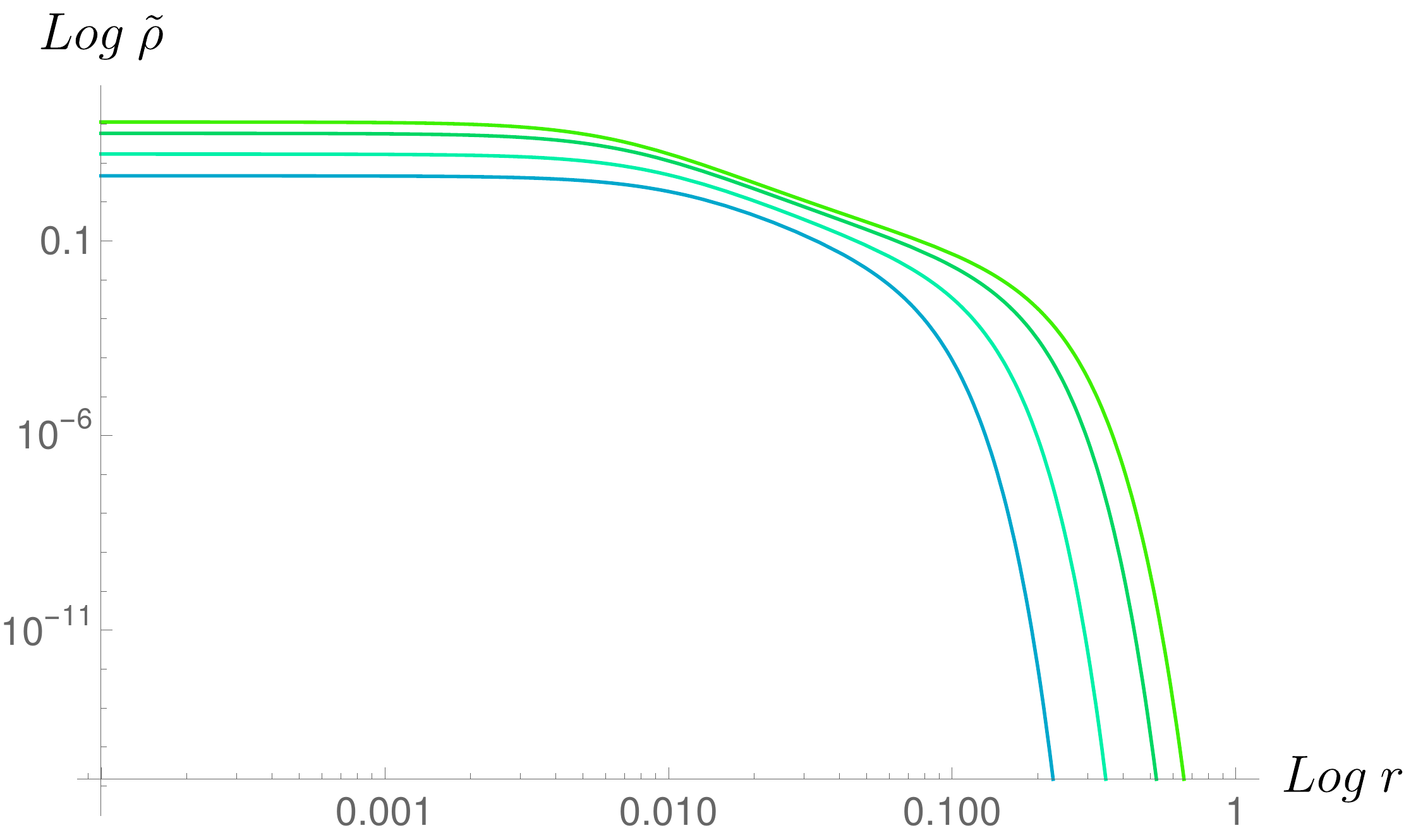}
\hfill
\includegraphics[width=0.45\textwidth]{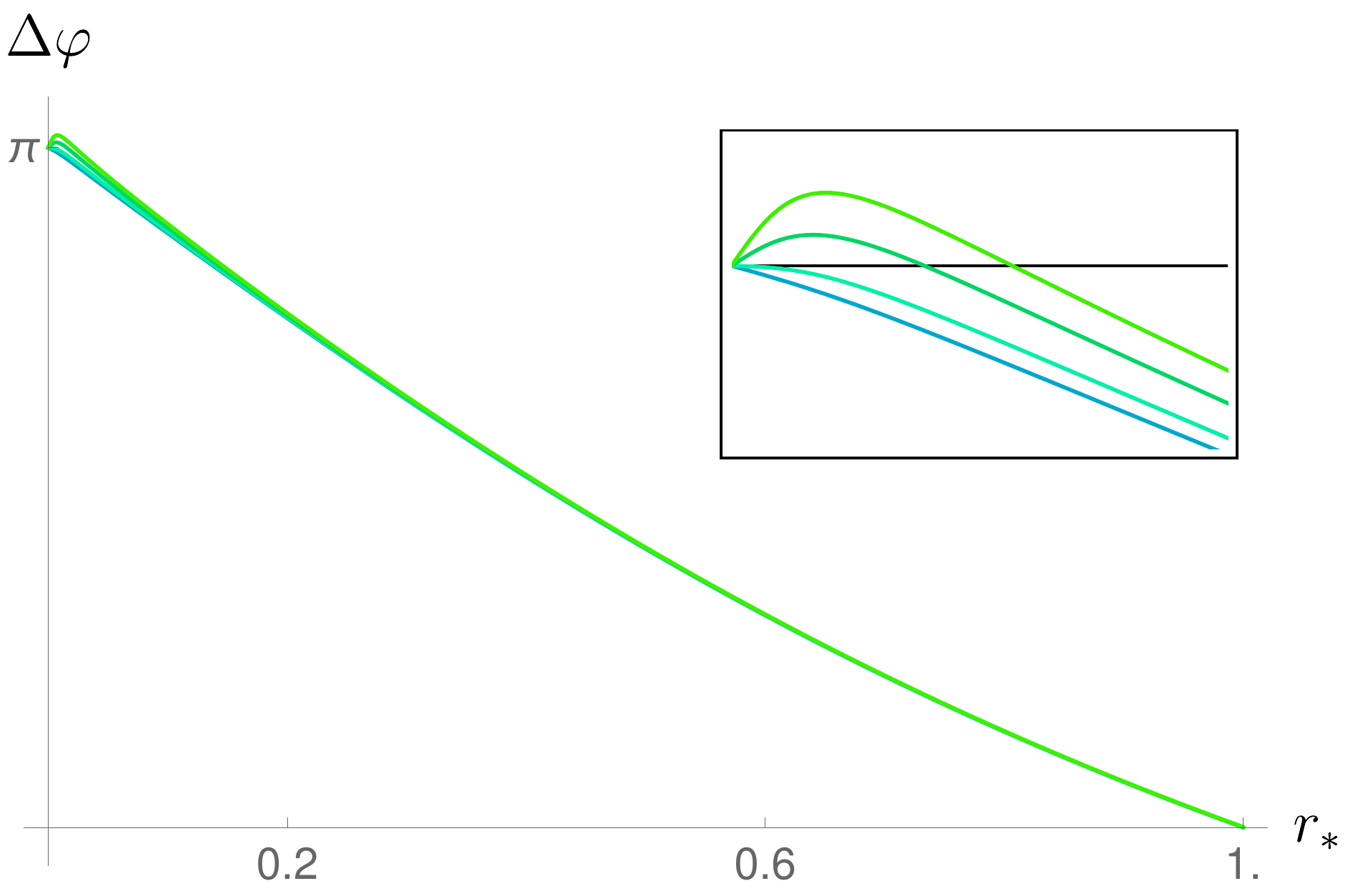}
\hfill
\\
\includegraphics[width=0.45\textwidth]{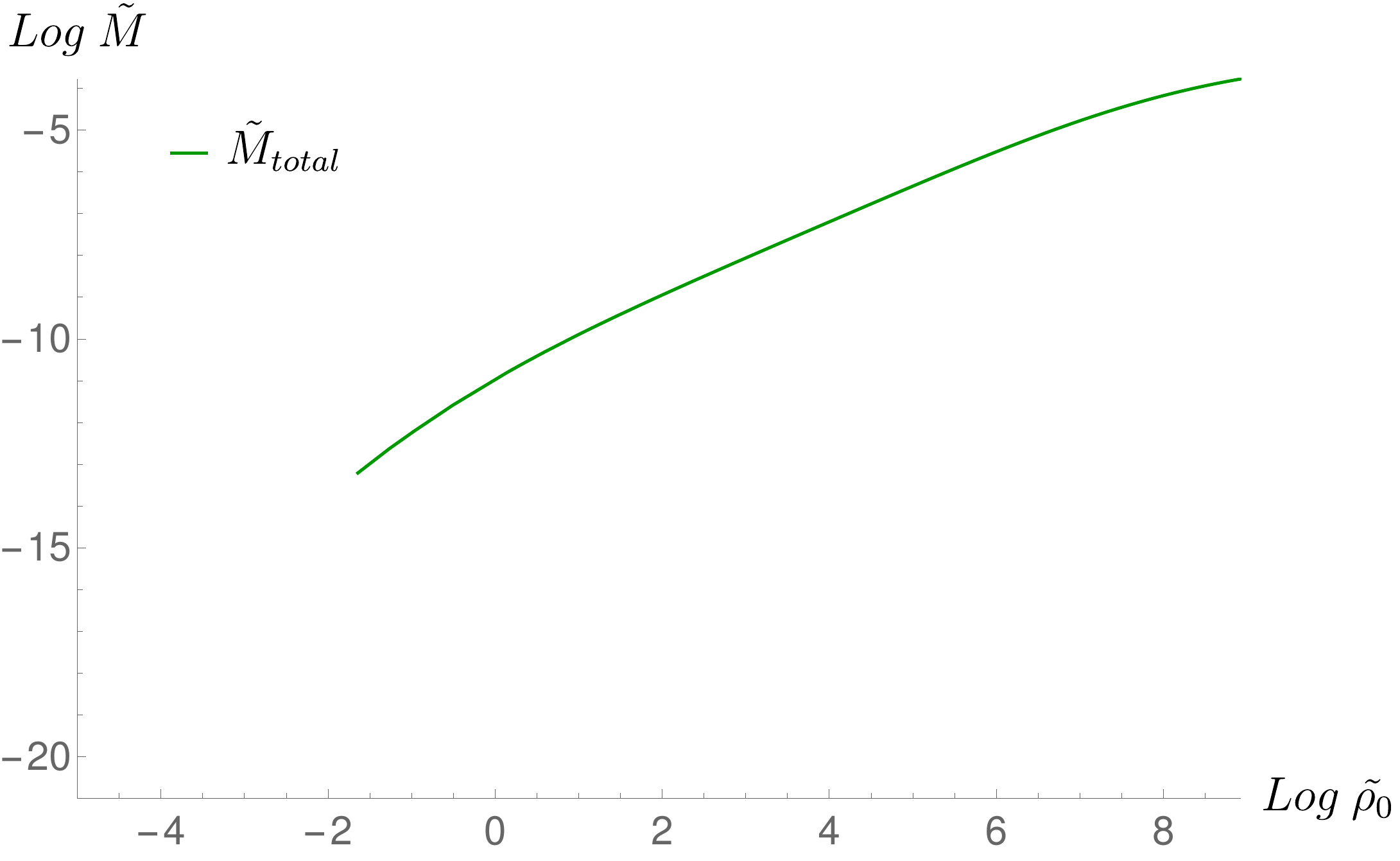}
\hfill
\includegraphics[trim=-5mm 0mm 0 10mm,width=0.2\textwidth]{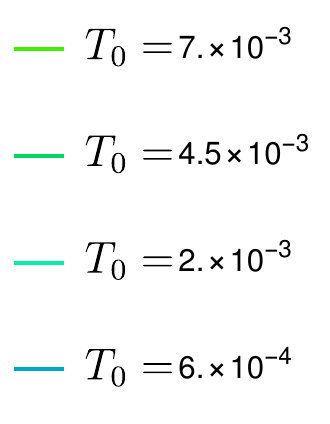}	
\put(-180,30){\scriptsize$\Theta_0=-15$\normalsize}
\put(-180,10){\scriptsize$\gamma=7\times10^5$\normalsize}
\put(-60,90){$\tilde{\phantom T}$}
\put(-60,64){$\tilde{\phantom T}$}
\put(-60,37){$\tilde{\phantom T}$}
\put(-60,11){$\tilde{\phantom T}$}
\caption{\label{perfilm10} Plots of the solutions corresponding to $\Theta_0=-10$ (first set of three plots) and $\Theta_0=-15$ (second set of three plots)  for different values of $\tilde{T}_0$.
\underline{Up-left:} logarithmic plot of the density profile as a function of the radius. The density has a plateau and decreases sharply at the boundary of the star for low temperatures, while for higher temperatures it takes a power law form. \underline{Up-right:} the angle $\Delta\varphi$ between the incident and scattered direction of a massive Euclidean geodesic. For the profiles with a power law decay, it grows into values bigger than $\pi$.
\underline{Down-left:} the total mass $\tilde M$ as a function of the central density. We see that there is no special feature at the temperatures at which the power law behavior of the star edge and the non-monotonic $\Delta\varphi$ show up, the maximum of the mass appearing at much higher temperatures.
}
\label{fig:negativetheta3}
\end{figure}
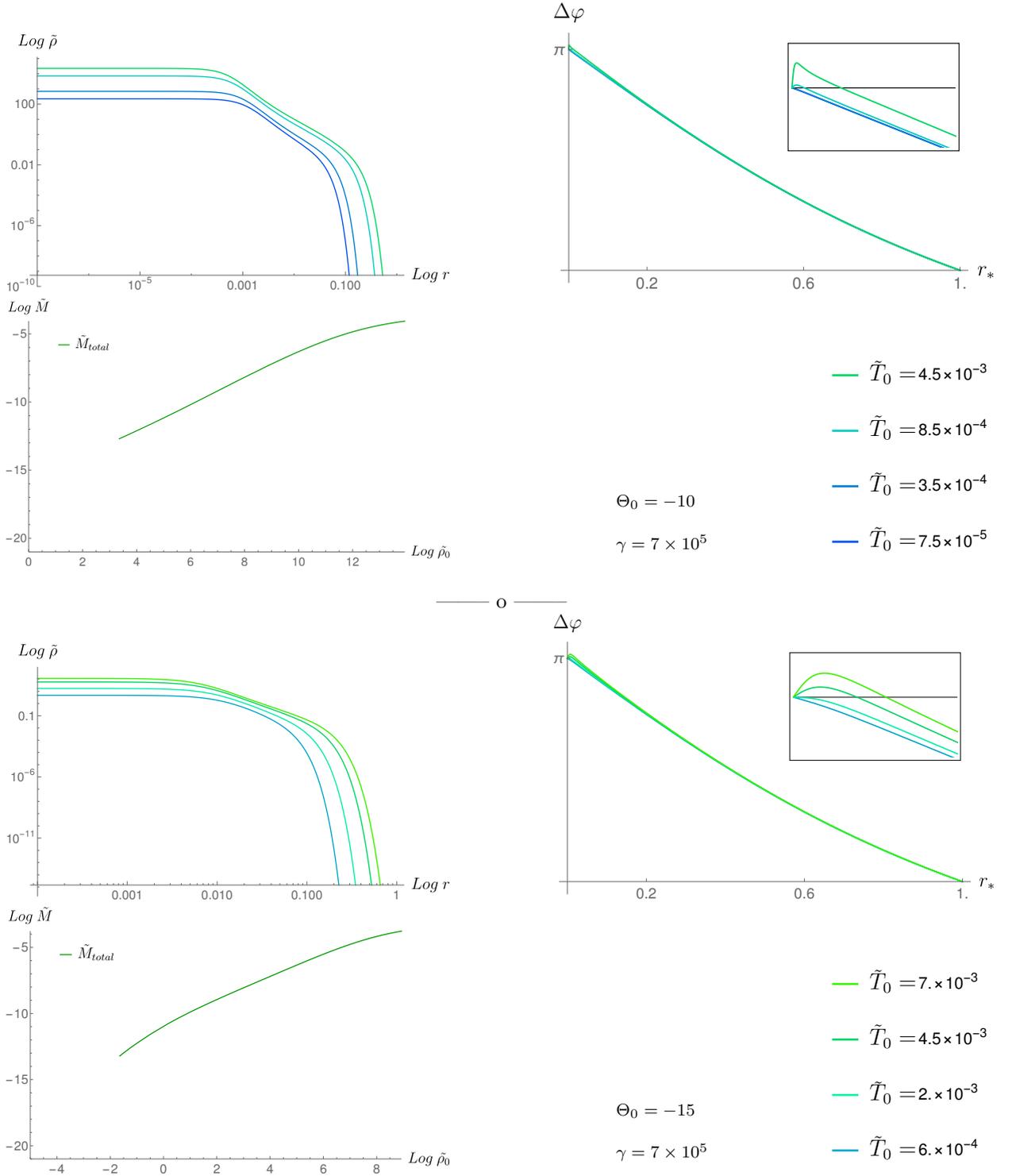

We solved numerically the above system of equations \eqref{eq:M1}-\eqref{eq:tolman1} with boundary conditions \eqref{eq:boundary.mu1} by using a {\tt Mathematica} routine. We explored the $(\tilde T_0,\Theta_0)$ plane for a fixed value of $\gamma^2$. At each point, we evaluated the function \eqref{eq:delta.varphi1} determining whether the angle $\Delta\varphi$ is a monotonic function of $r_*$. The results are shown in Figs. \ref{fig:positivetheta1} and \ref{fig:negativetheta3}.

At positive enough central degeneracy $\Theta_0 \gtrapprox 10$ and sufficiently small central temperature $\tilde T_0$, the density profiles have a well defined core-halo structure as exemplified in Fig. \ref{fig:positivetheta1}, in complete analogy with the asymptotically flat space \cite{2006IJMPB..20.3113C, 2015PhRvD..92l3527C, 2019arXiv190810303C, 2018arXiv180801007A,  2015MNRAS451622R,2018PDU....21...82A}. For such range of parameters, the highly-dense core is supported by degeneracy pressure while the outer halo (or diluted atmosphere) is hold against gravity by thermal pressure. The halo ends at a sharp edge at which the boundary of the star is reached, thanks to the overall AdS inward-pressure term. As the central temperature is increased at fixed central degeneracy, the compacity of the central core increases, until the critical temperature  $\tilde T^{cr}_0$ of the turning point is reached. Interestingly, above such critical value, {\em  e.g.} $\tilde T^{cr}_0 \approx 10^{-2}$ for $\Theta_0 = 30$, the outer halo region (just before the sharp edge) as well as the inner dense-core develop a power law morphology analogously as found in the asymptotically flat case \cite{1990A&A...235....1G}. The presence of a gravitational collapse to a singularity is evidenced through the sufficient condition of a local maximum of the total mass as a function of the central density (see the examples of $\Theta_0 = 30, 50$ in Fig. \ref{fig:positivetheta1}) \cite{Schiffrin:2013zta}. Its existence is typical of any relativistic self-gravitating system of fermions, as shown  in \cite{Arsiwalla2010} within the fully degenerate limit ($\tilde T \rightarrow 0$) in AdS, and in \cite{2014IJMPD..2342020A} or \cite{2018JHEP...05..118A} for the more general finite-temperature cases either in flat space, or AdS respectively. Remarkably, by increasing further the central temperature, a non-monotonic behaviour on the Euclidean geodesics appears (see Fig. \ref{fig:positivetheta1}). This result was not known from the recent related work \cite{2018JHEP...05..118A} by the authors, since only $\tilde T_0 \lesssim \tilde T^{cr}_0$ values were there explored.

At negative central degeneracy $\Theta_0 < 0$, the fermions are in the Boltzmannian regime of the Fermi-Dirac distribution, thus the star is fully supported by thermal pressure (for fixed AdS length). In this case the profiles have no core, but just an overall diluted density profile with an inner constant value followed by a smooth transition towards the sharp edge, for small enough central temperature $\tilde T_0$, see Fig. \ref{fig:negativetheta3}. As the central temperature is increased, the star gets more extended and massive, with an outer region, just before the sharp edge, developing a power law trend with a mild (non degeneracy-supported) core at the center. Such morphological density profile behavior resembles somehow the $\Theta_0 > 10$ case, and imply a non-monotonic trend of the geodesics. At some point, $\tilde T_0$ would reach the critical temperature for gravitational collapse, though such value is above our maximum parameter-space coverage {\em e.g.} $\tilde T_0 < 10^{-1}$, and certainly close to the ultra relativistic regime $\tilde T_0 \sim 1$. Consequently, in the total mass {\em vs.} central density plots the maximum is not reached, see Fig. \ref{fig:negativetheta3}.

These results can be summarized in a phase diagram as shown in Fig. \ref{fig:transitionregion}. The power law behavior at the boundary of the density profile, as well as the non-monotonic behavior on the geodesics, appear at intermediate degeneracies and high enough temperatures. Remarkably, this can be linked with the holographic perspective physics, as explicited in section \ref{sec:boundary} below.

The fact that $\tilde T^{cr}_0$ is considerably larger for negative to small central degeneracies respect to the $\Theta_0 > 10$ case discussed above, is understood in terms of the maximum possible pressure holding the star against gravity. While in the first scenario the fermions needs to be close to the ultra-relativistic regime for the thermal pressure to acquire its maximum value, in the later, is the fully degeneracy limit, given by $\Theta_0\gg 1$ for low enough $\tilde T_0$, the one halting gravity before the collapse. Moreover, it can be shown that while the critical mass for collapse is ${\rm Log} \tilde M^{cr} \approx -6$ in the highly degenerate cases, see Fig. \ref{fig:positivetheta1}, it starts to grow above this value for lower and lower $\Theta_0$ showing the temperature effects, in complete analogy to asymptotically flat case \cite{2014IJMPD..2342020A}. Finally, for positive and small central degeneracies $0 \lesssim \Theta_0 \lesssim 10$, the fermions are in a semi-degenerate regime developing a mild core - diluted halo behaviour, though the core does not fulfill the  quantum degeneracy condition $\lambda_B > 3 l$, with $\lambda_B$ the thermal de-Broglie wavelength and $l$ the interparticle mean distance, as first demonstrated in \cite{2015MNRAS451622R} in flat space. 

\newpage


\section{The holographic perspective}
\label{sec:boundary}
From the holographic point of view, the system under study corresponds to a degenerate fermionic state on a conformal field theory defined on a sphere. The boundary value of the local chemical potential acts as a source for the particle number operator in the boundary theory. It can be thus identified with the boundary chemical potential. On the other hand, the bulk temperature sets the Euclidean length of the thermal circle. When approaching the boundary, such length defines the boundary temperature. In consequence, the holographic CFT state is defined at finite chemical potential \cite{DeBoer2009} and finite temperature \cite{2018JHEP...05..118A}. For that reason, in this section we study the corresponding thermodynamics in the grand canonical ensemble \cite{new1}. 

To study the thermodynamic stability, the Katz criterion was used in \cite{2006IJMPB..20.3113C, 2015PhRvD..92l3527C} in the framework of self gravitating fermions in Newtonian gravity, and in \cite{2019arXiv190810303C, 2018arXiv180801007A} for the general relativistic case. Here we employ it in the asymptotically AdS holographic setup.
\subsection{Building the boundary state: the grand canonical potential}
\label{subsec:canonicalpotential1}
\paragraph{The on shell action: }
\label{sec:potential}
The grand canonical potential $\Omega(\mu,T)$ is calculated according to
\begin{equation}
\Omega(\tilde\mu,\tilde T)  = m\tilde T\, S_{\sf on-shell}^E
\,,
\label{eq:grand.canonical.potential}
\end{equation}
where $\tilde T$ and $\tilde \mu$ are the dimensionless boundary temperature and chemical potential respectively, obtained as the asymptotic values of the corresponding bulk functions. The magnitude $S_{\sf on-shell}^E$ is the Euclidean action obtained by a Wick rotation $t_{E}=it$ of the gravitational and perfect fluid action (see Appendix \ref{sec:background}), supplemented with the Gibbons-Hawking-York term and the necessary holographic renormalization counterterms
\begin{equation}
S^E_{\sf on-shell}=\left.\left( S_{\sf Gravity}^E+S_{\sf Fluid}^E+S_{\sf GHY}^E+S_{\sf ct}^E\right)\right|_{\sf on-shell}
\,.
\end{equation}
By evaluating these terms on the solution, we get
\begin{align}
\Omega(\tilde\mu,\tilde T)
&=
\frac{L }{G\, }e^{-\frac{\nu_\infty}{2}}
\left(
\tilde M-4\pi \int dr
\,e^{\frac{\nu+\lambda}{2}}r^2
(\tilde\rho+\tilde P)
\right)\,,
\label{eq:on.shell.action}
\end{align}
where $\tilde M$ and $\nu_\infty$ correspond to the asymptotic value of the $\tilde{M}(r)$ and $\nu(r)$ respectively.

The on shell action is a natural function of the central parameters $\tilde T_0$ and $\Theta_0$. Since we are interested in the boundary physics, we want to write it as a parametric function of the boundary quantities $\tilde T$ and $\tilde \mu$, which depend on $\tilde T_0$ and $\tilde \mu_0$ according to \eqref{eq:tolman1}. Since the dependence is parametric, there is a chance that $\Omega(\tilde T,\tilde \mu)$ is not single-valued. This situation is somewhat frequent in self gravitating systems, and we will discuss the issue in more detail below.

\paragraph{Stability analysis:}
\label{sec:stability}
Since the grand canonical potential might be multi-valued, we need a criterion to determine which of its many branches corresponds to a stable phase of the boundary theory at a given temperature $\tilde T$ and chemical potential $\tilde \mu$. We use the Katz stability criterion, that we sketch below.

We start with the entropy $S$ and define the grand canonical free entropy as
\begin{equation}
\Phi(\tilde{T},\tilde{\mu})=
S - \frac{1}{\tilde{T}}\tilde{M}-\frac{\tilde{\mu}}{\tilde{T}}N
=-\frac{\Omega}{\tilde T}
\,,
\label{eq:free.entropy}
\end{equation}
such that its derivative with respect to $1/\tilde{T}$ gives minus the mass $-\tilde{M}$ of the configuration. Next we assume that $\Phi(\tilde T,\tilde \mu)$ can be extended away from the equilibrium configuration into a function $\Phi^{\sf ext}(q_i,\tilde T,\tilde \mu)$ that depends on the generalized coordinates $q_i$ that parameterize the deviation from equilibrium.
A simple example of such generalized coordinates can be given in the non-gravitational case, in which the equilibrium state is homogeneous. Any local thermodynamic quantity, such as the temperature or chemical potential, can then be decomposed into Fourier modes. Thus, aside from the constant contribution, the amplitude of any higher mode parameterizes the deviation from homogeneity and can be identified with one of the $q_i$. In the self-gravitating case, the equilibrium state is given by the inhomogeneous solution obtained in section \ref{sec:background1}. Any deformation of the corresponding local thermodynamic quantities moves the system out of equilibrium. By decomposing these deformations in a suitable basis, we get the generalized coordinates $q_i$. In any case, for the derivation below an explicit identification of the $q_i$ is not needed, being enough to acknowledge that  they exist. The equilibrium states are stationary points of the extended grand canonical free entropy $\Phi^{\sf ext}$.
\begin{equation}
 \partial_i\Phi^{\sf ext}(q_i,\tilde T,\tilde \mu) = 0\,.
\label{eq:equilibrium.condition}
\end{equation}
We can write the equilibrium solutions of this equation as $q_i=q_i(\tilde T,\tilde\mu)$, and recover the equilibrium grand canonical free entropy as
\begin{equation}
\Phi(\tilde T,\tilde \mu)=\Phi^{\sf ext}(q_i(\tilde T,\tilde \mu),\tilde T,\tilde \mu).
\label{eq:extended.free.entropy}
\end{equation}

Since the derivative of $\Phi(\tilde{T},\tilde{\mu})$ with respect to the inverse temperature $1/\tilde{T}$ at fixed $\tilde{\mu}/\tilde{T}$ gives minus the mass $-\tilde{M}$, we can define an extended mass function as $-\tilde{M}^{\sf ext}=\partial_{1/\tilde{T}}\Phi^{\sf ext}(q_i,\tilde{T},\tilde{\mu})$. Using \eqref{eq:equilibrium.condition} and \eqref{eq:extended.free.entropy}
we can show that the equilibrium mass reads
\begin{equation}
-\tilde{M}(\tilde{T},\tilde{\mu})=
-\tilde{M}^{\sf ext}\left(q_i(\tilde T,\tilde \mu),\tilde T,\tilde \mu\right)\,.
\end{equation}
A derivative of this equation then gives
\begin{equation}
-\partial_{1/\tilde T}\tilde{M}=
-\partial_{1/\tilde T}\tilde{M}^{\sf ext}
-\sum_i
\partial_i\tilde{M}^{\sf ext} \,\partial_{1/\tilde T}q_i\,.
\label{eq:mass.derivative}
\end{equation}
The second term can be rearranged by using the implicit function theorem in \eqref{eq:equilibrium.condition}, to get
\begin{equation}
-\partial_{1/\tilde T}\tilde{M}=
-\partial_{1/\tilde T}\tilde{M}^{\sf ext}
-\sum_{ij}
\partial_iM^{\sf ext} \,(\partial_i\partial_j\Phi^{\sf ext})^{-1}\partial_j \tilde{M}^{\sf ext} \,.
\label{eq:kats.criterion}
\end{equation}
We can parameterize the deformation away from equilibrium with coordinates $q_i$ such that the matrix $\partial_i\partial_j\Phi^{\sf ext}$ is diagonal, as
\begin{equation}
-\partial_{1/\tilde T}\tilde{M}=
-\partial_{1/\tilde T}\tilde{M}^{\sf ext}
-\sum_i
\frac{(\partial_i \tilde{M}^{\sf ext} )^2}{\lambda_i}\,.
\label{eq:kats.criterion.diagonal}
\end{equation}
where $q_i$ is now the direction in the deformation space characterized by the eigenvalue $\lambda_i$ of the matrix $\partial_i\partial_j\Phi^{\sf ext}$. When any of such eigenvalues, say $\lambda$, is close enough to zero, it dominates the right hand side of equation \eqref{eq:kats.criterion.diagonal}, resulting in
\begin{equation}
-\partial_{1/\tilde T}\tilde{M}\approx
-\frac{(\partial_{q}\tilde{M}^{\sf ext})^2}{\lambda}\,,
\end{equation}
The crucial observation is that the sign of the eigenvalue is opposite to the sign of the expression $-\partial_{1/\tilde T}\tilde M$. Since this is valid whenever $\lambda$ is approaching zero, the derivative is diverging at such points. In conclusion, whenever the plot of $-\tilde{M}$ versus $1/\tilde{T}$ at constant $\tilde{\mu}/\tilde{T}$ has a vertical asymptota, the slope of the curve at each side of the asymptota is opposite to the sign of the eigenvalue of $\partial_i\partial_j\Phi^{\sf ext}$ that is approaching zero at that temperature.

In a stable or meta-stable equilibrium state, the grand canonical free entropy $\Phi^{\sf ext}$ is a maximum. This implies that all the eigenvalues of $\partial_i\partial_j\Phi^{\sf ext}$ are negative. As we move the inverse temperature at fixed $\tilde{\mu}/\tilde{T}$, the system evolves and the plot $-\tilde{M}$ versus $1/\tilde{T}$ eventually reaches a vertical asymptota. Since all the eigenvalues are negative, it approaches it with a positive slope. If at the other side of the asymptota the slope becomes negative, this implies that one of the eigenvalues changed its sign, and  the system reached an unstable region.

In what follows we identify the stable equilibrium state in which all the eigenvalues are negative with the diluted configurations, for which the central degeneracy $\Theta_0$ is negative. Next, we follow the $-\tilde{M}$ versus $1/\tilde{T}$ curve until we reach an asymptota at which the slope changes its sign. For each change from positive to negative slope, we count a new positive eigenvalue. For each change from negative to positive slope, we count a new negative eigenvalue. Any region with at least one positive eigenvalue, is unstable.

\newpage 
\subsection{Probing the boundary state: the correlator of a scalar operator}
\label{sec:correlators1}
\vspace{-.15cm}
The spacelike geodesics we calculated in Section \ref{subsec:geodesics1} can be obtained as the trajectories of a particle of mass ${\sf m}$  with an Euclidean worldline, see Appendix \ref{sec:correlators}. The corresponding action, evaluated on shell, reads
\vspace{-.5cm}
\begin{equation}
S^{E~ \sf on-shell}_{\sf Particle}=2{\sf m}L\int_{r_{*}}^{r_{\epsilon}}dr
\frac{r e^{\frac{\lambda(r)}{2}}}{ \sqrt{r^2-r_{*}^2}}.
\label{eq:action.particle.on.shell1}
\end{equation}

Equation \eqref{eq:action.particle.on.shell1} gives the on-shell action as a function of the position of the tip of the geodesic $r_*$. Equation \eqref{eq:delta.varphi1} on the other hand, provides $\Delta\varphi$ as a function of the same variable. This allows us to parametrically plot the the on-shell action as a function of the angular separation. For the cases in which $\Delta\varphi$ is non-monotonic as a function of $r_*$, the Euclidean particle action becomes multi-valued.

Using the dictionary of the AdS/CFT correspondence \cite{bala, balabala, noss}, we can now evaluate the two-point correlator of a scalar operator in the limit of a large conformal dimension $\Delta\equiv {\sf m}L$, as the exponential of minus the absolute minimum of the Euclidean particle action. 
\subsection{Results from the holographic perspective}
\vspace{-.15cm}
We calculated the grand canonical free entropy $\Phi(\tilde T,\tilde{\mu})$ for our numerical solutions, as well as their total mass $\tilde M(\tilde T,\tilde{\mu})$. The resulting curves are shown in Figs. \ref{fig:massvstemp06} and \ref{fig:massvstemp062}, where we plot both magnitudes as functions of the inverse temperature at fixed $\tilde{\mu}/\tilde{T}$. They are multi-valued functions as expected from their parametric definitions. Starting from the most negative value of $\Theta_0$, identified as the stable branch, the curves of $-\tilde{M}$ {\em vs.} $1/\tilde{T}$ spirals into a succession of vertical asymptotae in which the slope changes from positive to negative. This indicates that a succession of eigenvalues are becoming positive, increasing the degree of instability. At a certain point the process reverses, and the curve spirals out. Nevertheless, the number of eigenvalues that change back to negative sign is not enough to restore local stability. At larger $\Theta_0$ the curve spirals again into the gravitational collapse. Thus from the holographic perspective, the instability can be identified as the onset of the confinement to deconfinement transition \cite{DeBoer2009, Arsiwalla2010} .

We also evaluated the Euclidean particle action \eqref{eq:action.particle.on.shell1}, whose negative exponential provides the two-point scalar correlator, for the different backgrounds. This is depicted on the left hand side of Fig. \ref{fig:correlator}. As the boundary temperature is increased, the action becomes multivalued, developing a ``swallow tail'' form. Since the scalar correlator is given by its smaller branch,  it develops a non-vanishing angular derivative at $\Delta\varphi=\pi$.  In Fig. \ref{fig:massvstemp06} the branches of the free entropy at which the correlator is multivalued are identified with a dotted line. Interestingly, the swallow tail always appears after the first eigenvalue has become positive, {\em i.e.} inside an unstable branch.

\section{Analysis of the results}
\label{sec:comparison}
\begin{figure}[H]
	\centering
	\includegraphics[width=.495\textwidth]{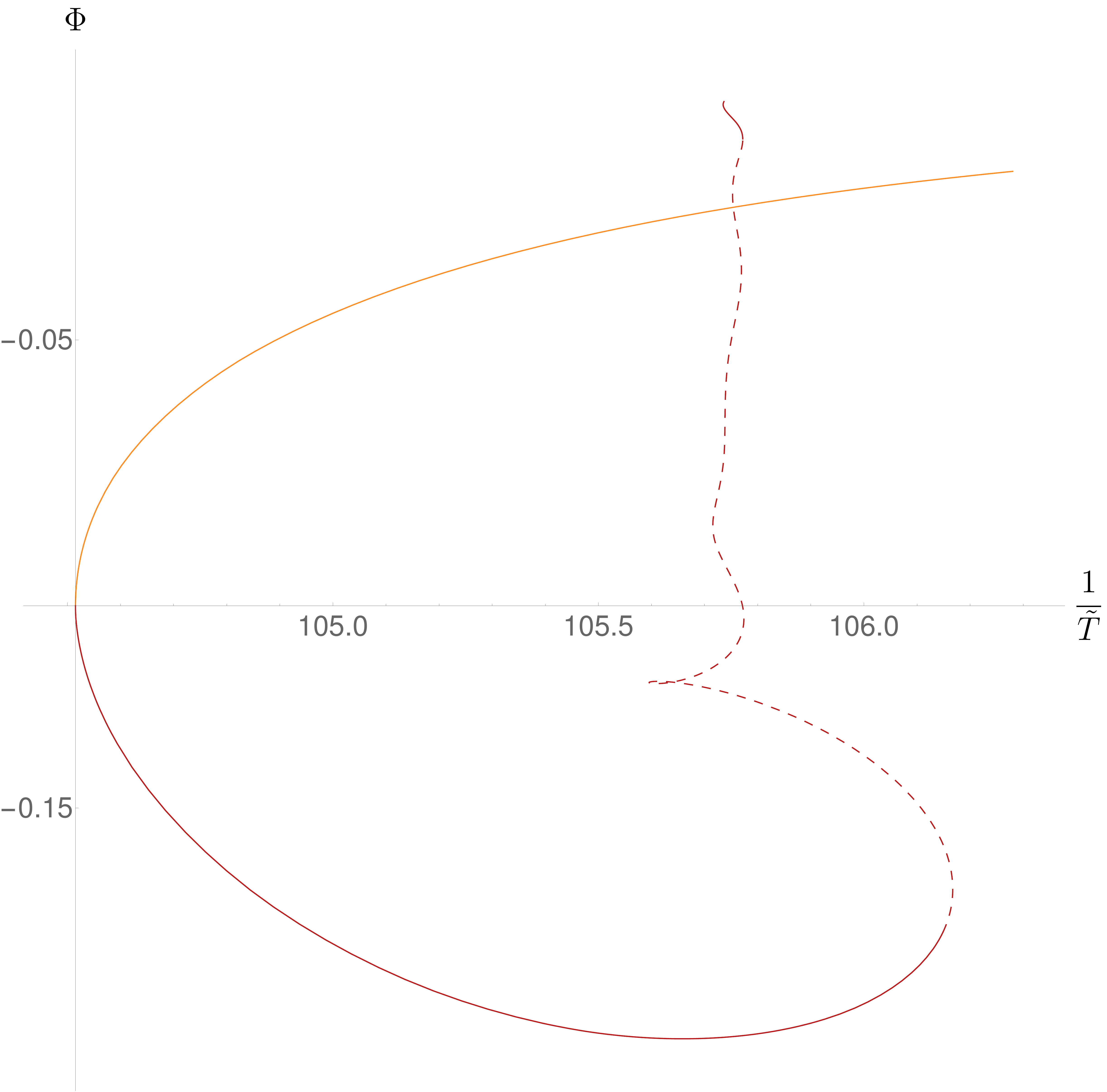}
	\hfill
	\includegraphics[width=.495\textwidth]{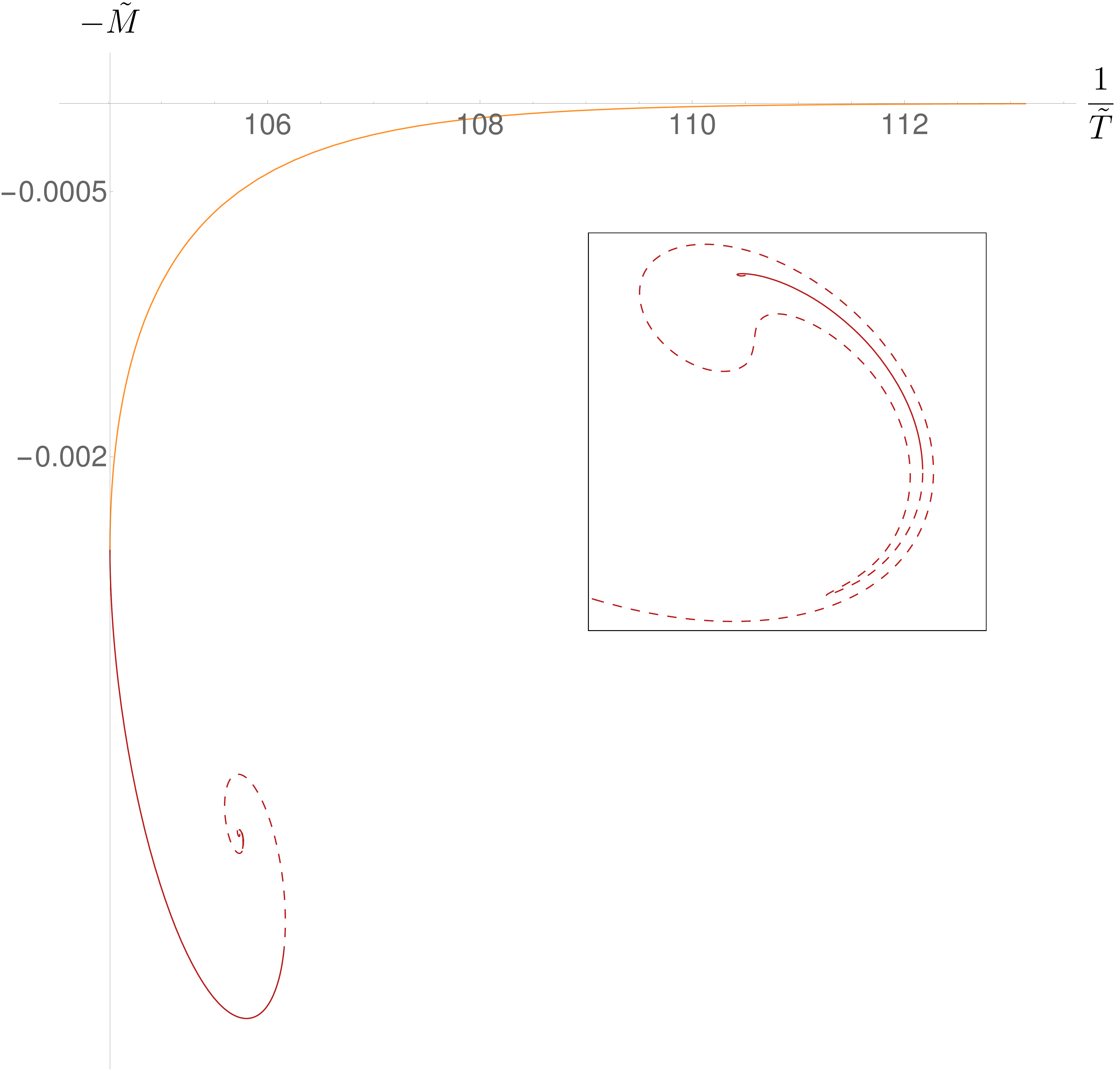}
\\
~
\\	
	\centering
	\includegraphics[width=.495\textwidth]{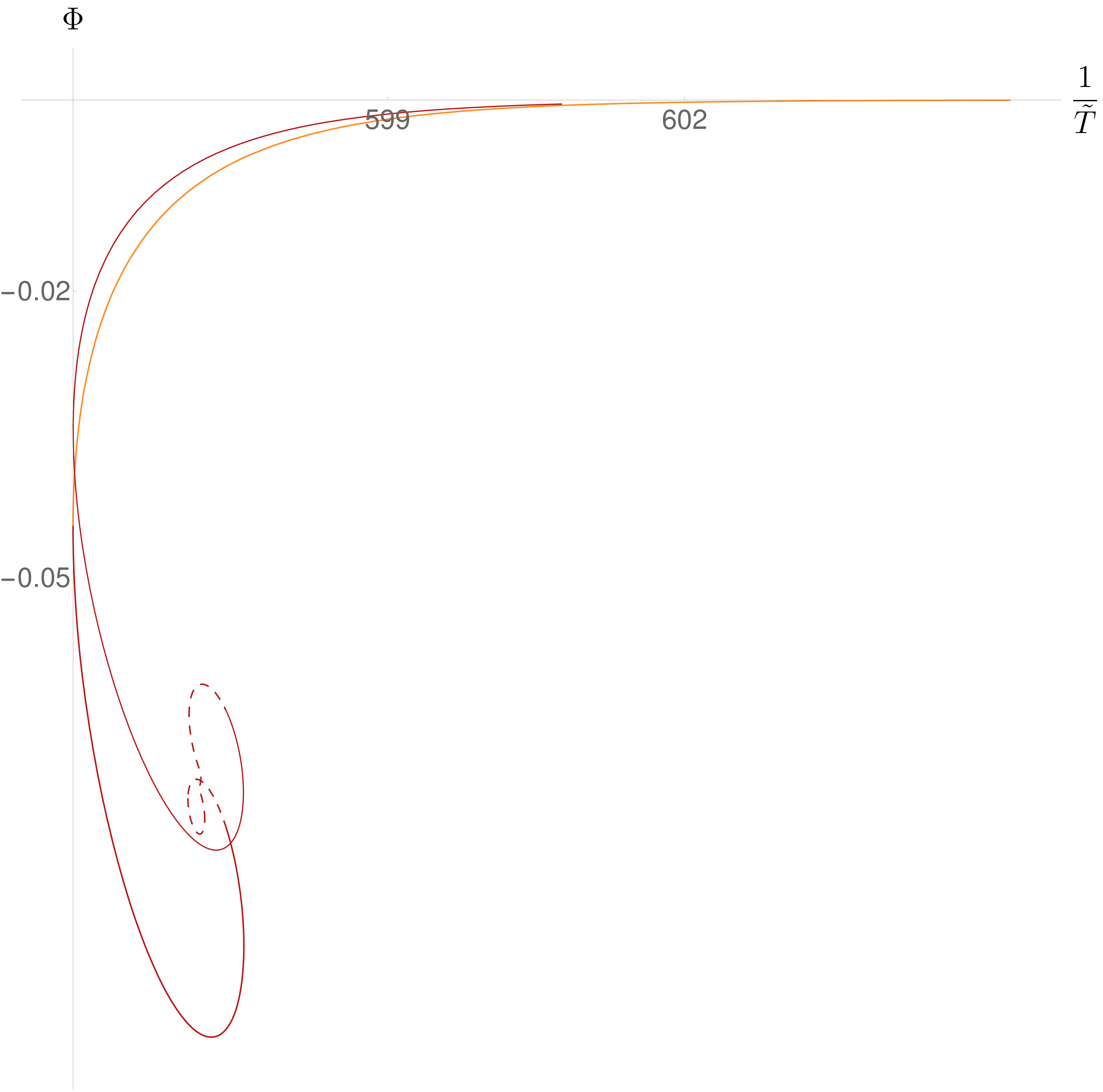}
	\hfill
	\includegraphics[width=.495\textwidth]{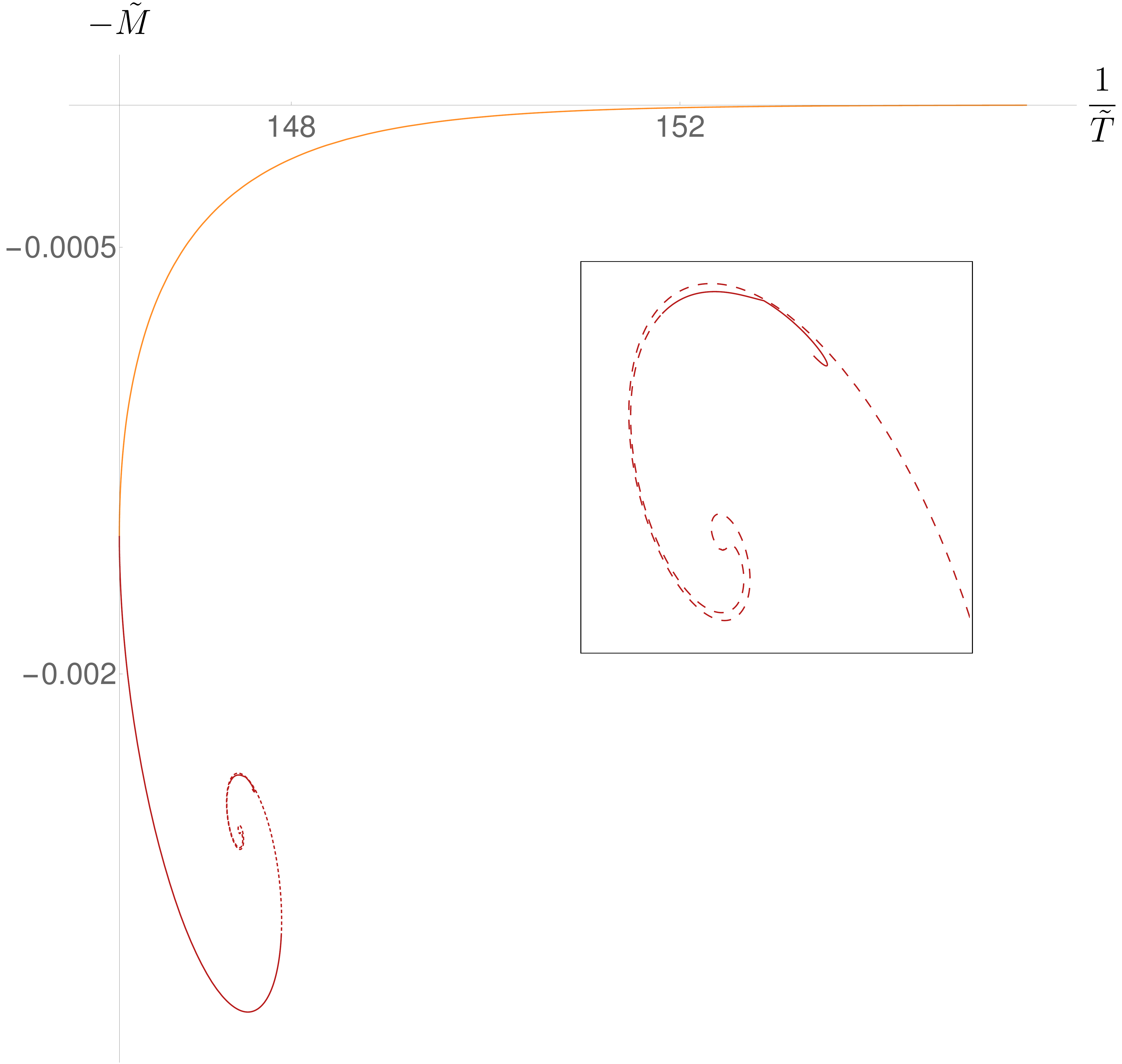}
	\put(-120,40){\scriptsize  \redlinedashed \hspace{0.2cm}Swallow Tail\normalsize}
	\put(-120,30){\scriptsize  \orangeline \hspace{0.2cm}Stable\normalsize}
	\put(-120,20){\scriptsize  \redline \hspace{0.2cm}Unstable\normalsize}
	\caption{\label{fig:massvstemp06} Parametric plot of the grand canonical free entropy $\Phi$ (left) and minus the total mass $-\tilde{M}$ (right) as functions of the inverse boundary temperature $1/\tilde{T}$, for a fixed value of $\mu/\tilde{T}=50$ (upper plots) and $\mu/\tilde{T}=80$ (bottom plots).
		The curve starts at negative values of $\Theta_0$ where all the eigenvalues are assumed to be negative (orange line). As the plot on right hand side reaches its first vertical asymptota, an eigenvalue changes sign from negative to positive (red line). The process repeats on each vertical asymptota in which the slope changes from positive to negative. At the center of the spiral (see inset) the process reverses and the slope starts to change from negative to positive at each vertical asymptota, changing back the sign of the corresponding eigenvalues. Nevertheless, the stability is not recovered.
The dashed line corresponds to the presence of a swallow tail on the scalar correlators, coinciding with a non-monotonic $\Delta\varphi$, and a power law edge in the density profiles. We see that such behaviour always appear in the unstable region.}
\end{figure}

\begin{figure}[H]
\centering
\includegraphics[width=.495\textwidth]{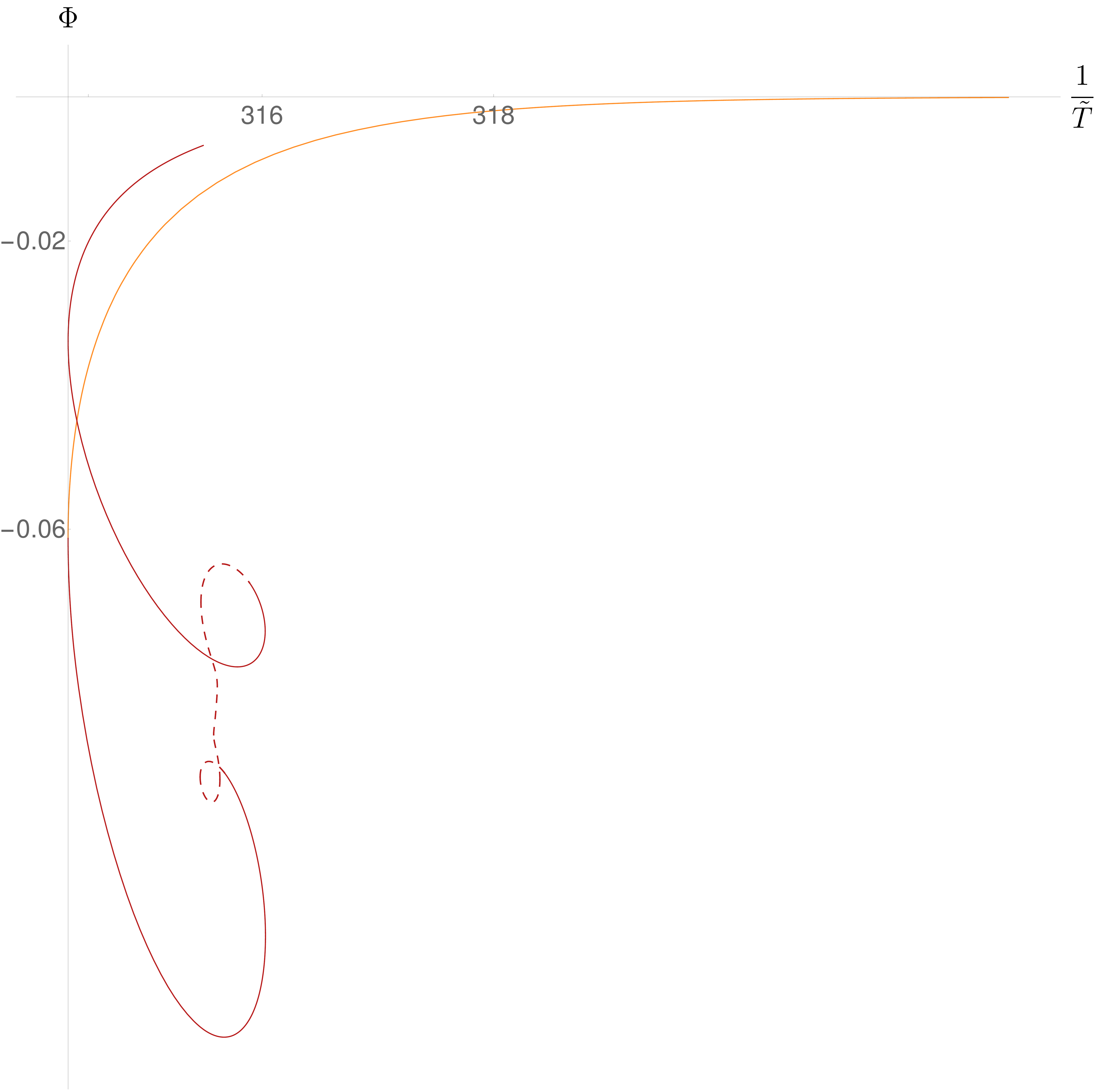}
\hfill
\includegraphics[width=.495\textwidth]{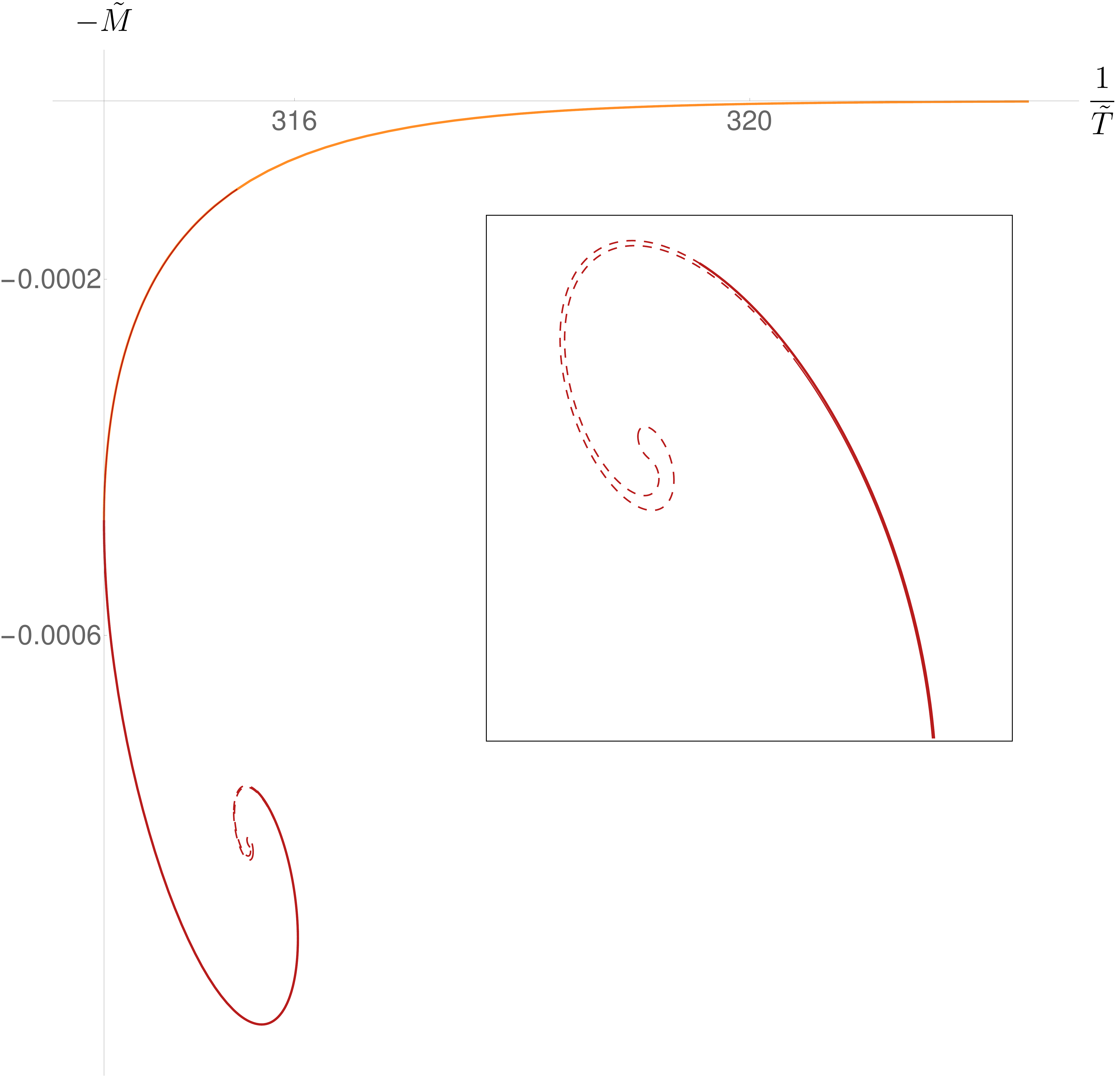}
\centering
\includegraphics[width=.495\textwidth]{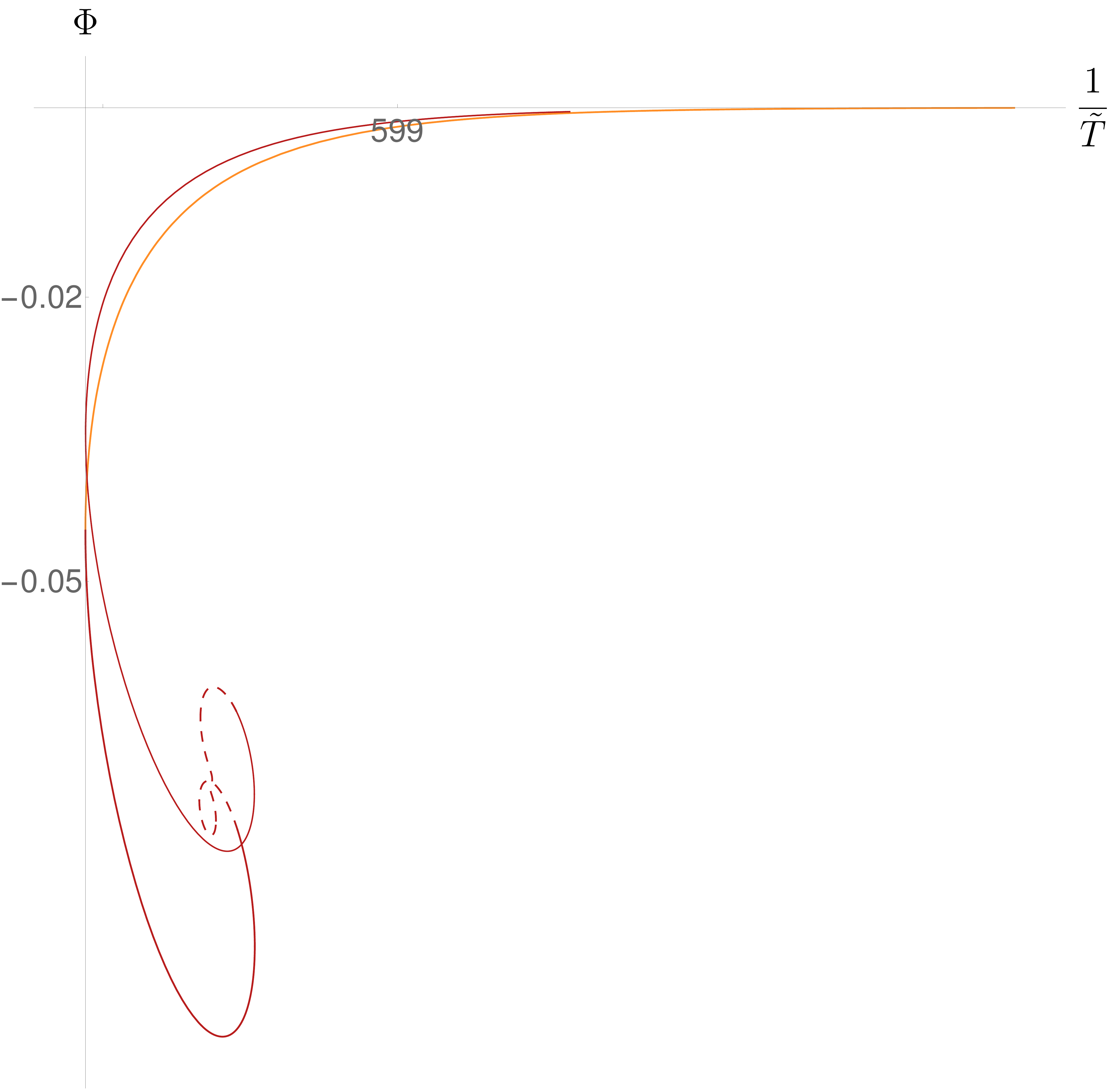}
\hfill
\includegraphics[width=.495\textwidth]{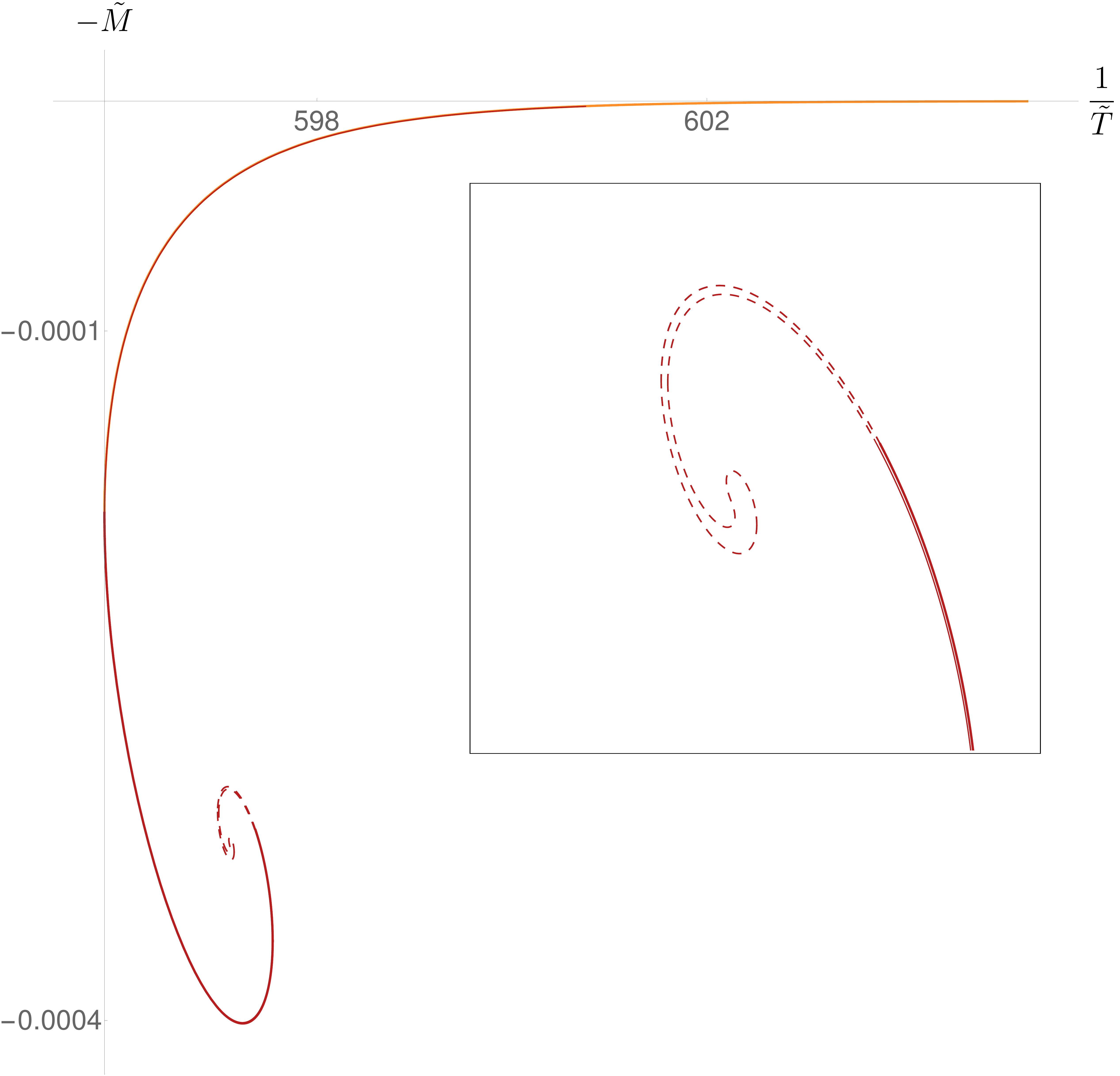}
\put(-120,40){\scriptsize  \redlinedashed \hspace{0.2cm}Swallow Tail\normalsize}
\put(-120,30){\scriptsize  \orangeline \hspace{0.2cm}Stable\normalsize}
\put(-120,20){\scriptsize  \redline \hspace{0.2cm}Unstable\normalsize}
\caption{\label{fig:massvstemp062} Parametric plot of the grand canonical free entropy $\Phi$ (left) and minus the total mass $-\tilde{M}$ (right) as functions of the inverse boundary temperature $1/\tilde{T}$, for a fixed value of $\mu/\tilde{T}=200$ (upper plots) and $\mu/\tilde{T}=400$ (bottom plots).
The curve starts at negative values of $\Theta_0$ where all the eigenvalues are assumed to be negative (orange line). As the plot on right hand side reaches its first vertical asymptota, an eigenvalue changes sign from negative to positive (red line). The process repeats on each vertical asymptota in which the slope changes from positive to negative. At the center of the spiral (see inset) the process reverses and the slope starts to change from negative to positive at each vertical asymptota, changing back the signs of the corresponding eigenvalues. Nevertheless, the stability is not recovered.
The dashed line corresponds to the presence of a swallow tail on the scalar correlators, coinciding with a non-monotonic $\Delta\varphi$, and a power law edge in the density profiles. We see that such behavior always appear in the unstable region.}
\end{figure}

\begin{figure}[H]
\centering
\includegraphics[width=0.45\textwidth, height=0.21\textheight]{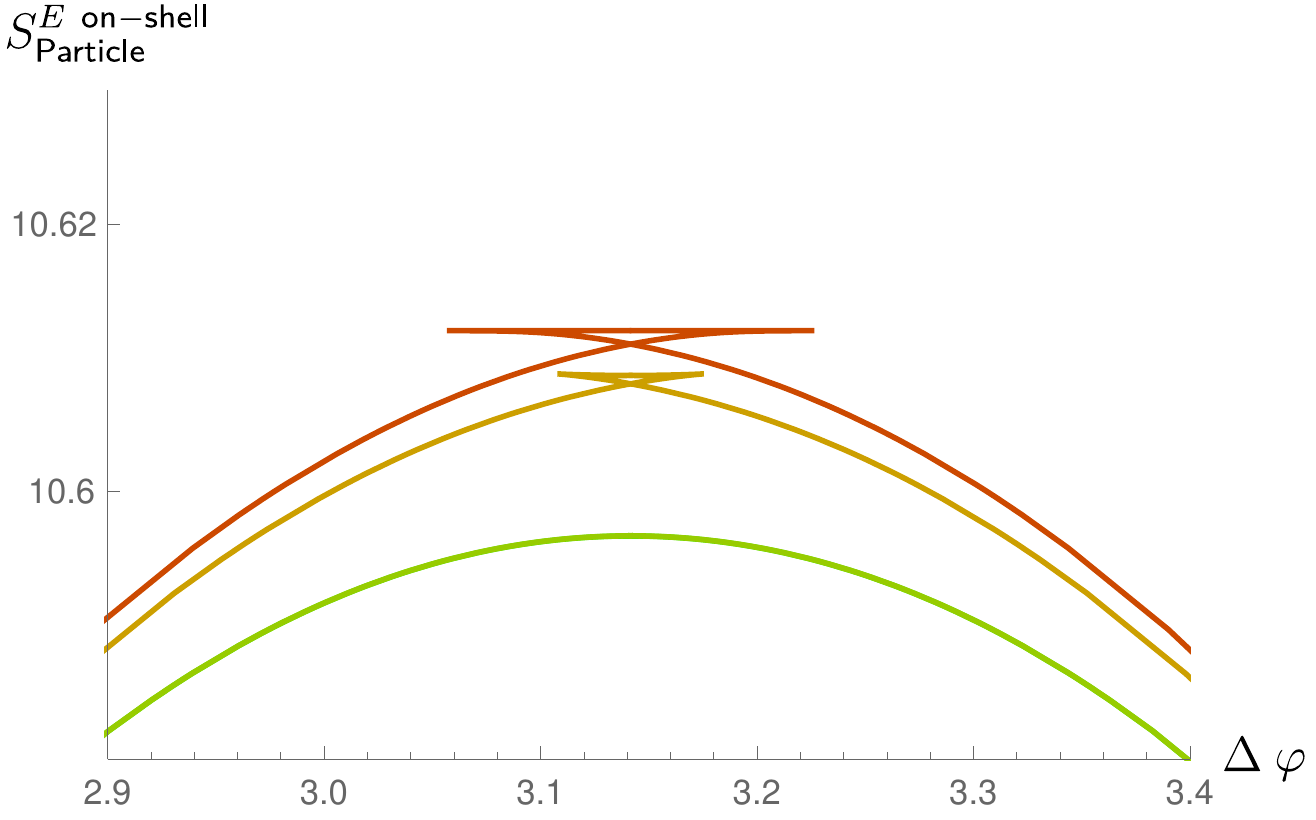}
\hfill
\includegraphics[width=0.4\textwidth]{all50.pdf}
\newline
\includegraphics[width=0.45\textwidth, height=0.20\textheight]{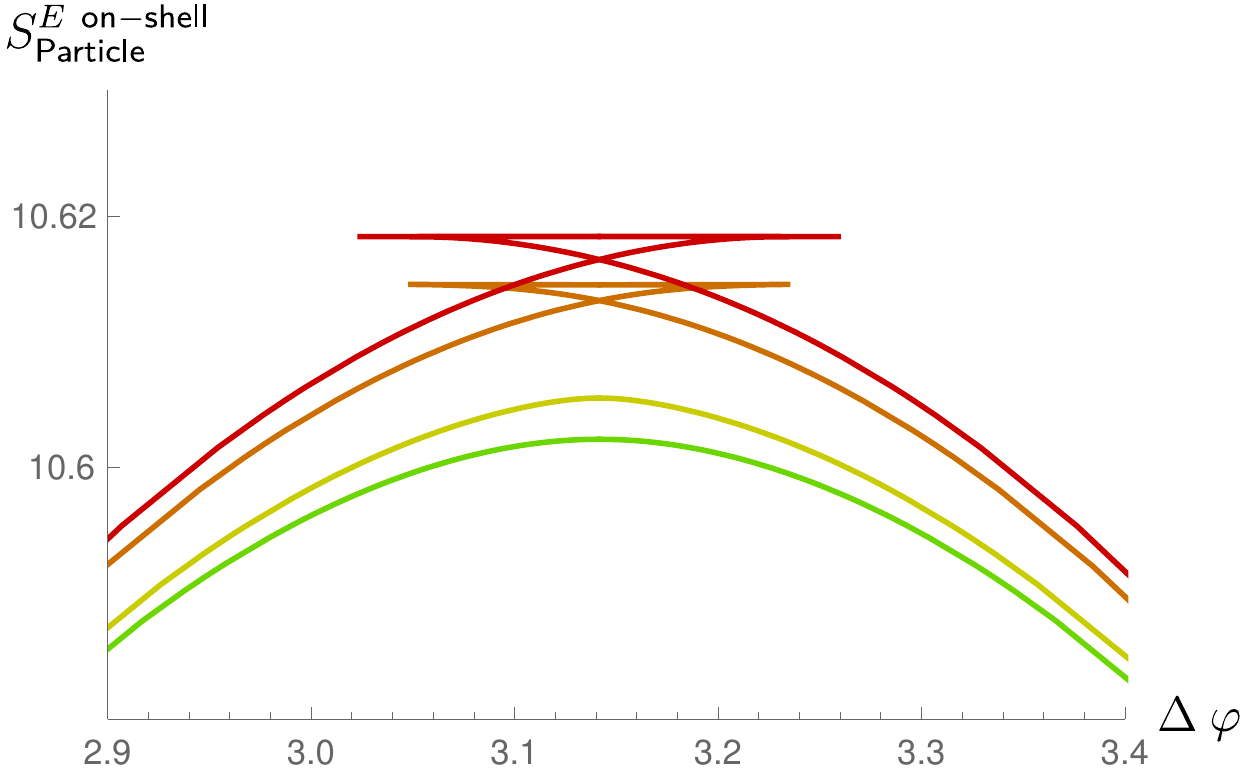}
\hfill
\includegraphics[width=0.4\textwidth]{all30.pdf}
\hfill
\\
\centerline{-------- o --------}
\includegraphics[width=0.5\textwidth]{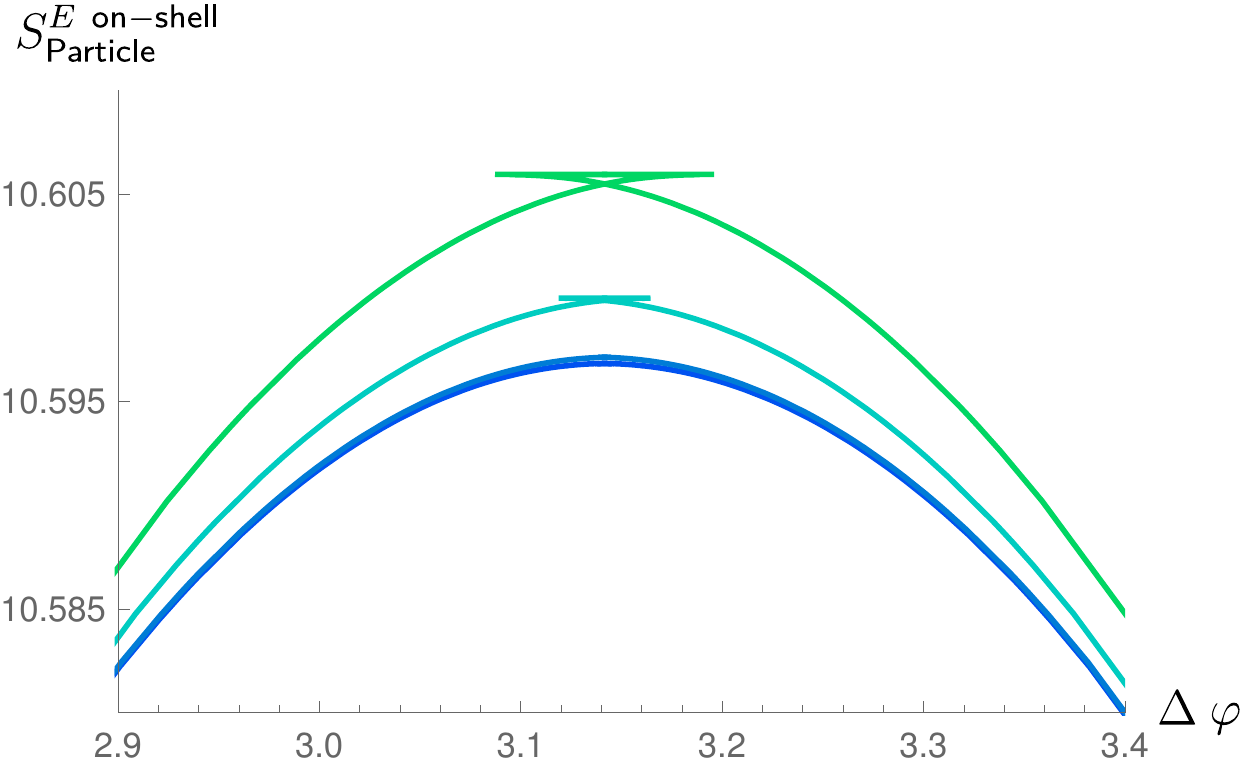}
\hfill
\includegraphics[width=0.4\textwidth]{allm10.pdf}
\includegraphics[width=0.5\textwidth]{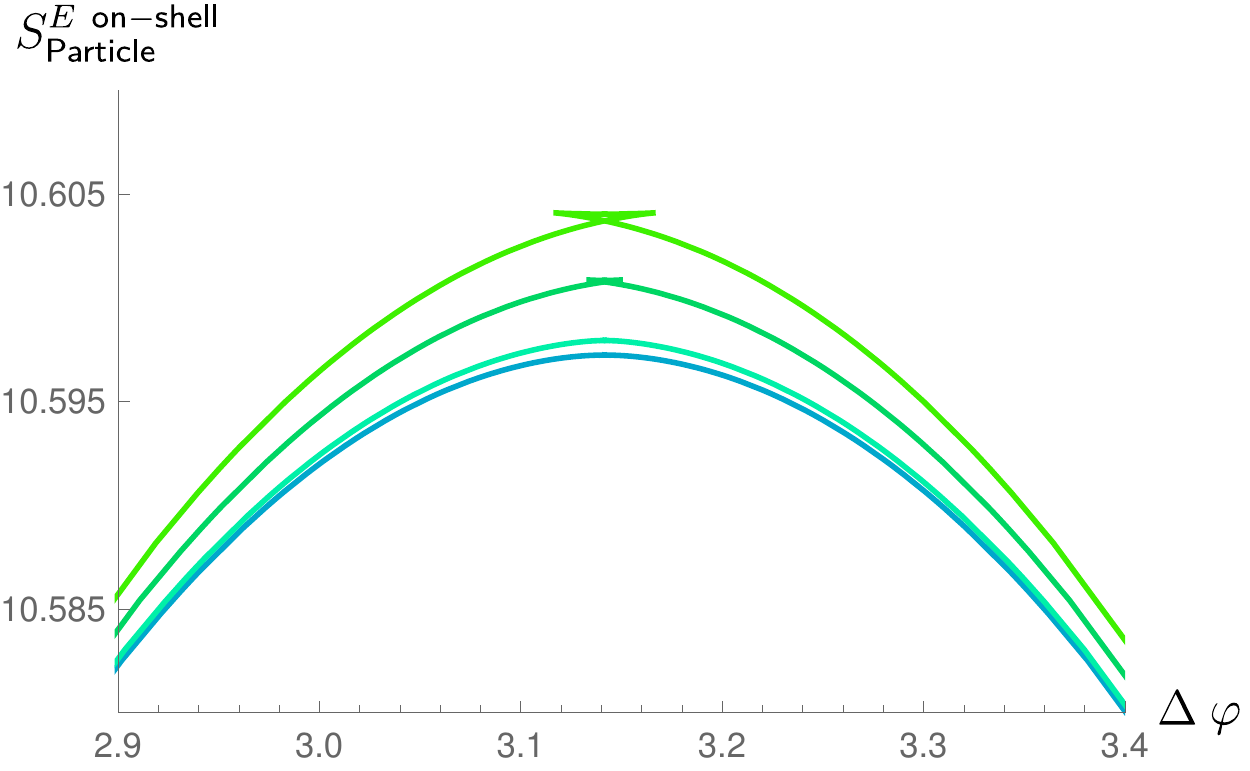}
\hfill
\includegraphics[width=0.4\textwidth]{allm15.pdf}
\caption{\label{fig:correlator} Plots corresponding to $\Theta_0=50$, $\Theta_0=30$, $\Theta_0=-10$ and $\Theta_0=-15$ (from top to bottom).
\underline{Left:} the on-shell Euclidean particle action as a function of the angular span $\Delta\varphi$ (the angular domain has been extended into negative values in order to get a clearer view); a swallow tail structure appears as the central temperature is increased. The scalar correlator corresponds to the smaller branch of such multivalued function, implying that it has a non-vanishing derivative at an angular separation of $\pi$. \underline{Right:} the angular span as a function of the geodesic tip $r_*$; the non-monotonicity is what originates the swallow tail on the left. Colors are correlated with Figs. \ref{fig:positivetheta1} and \ref{fig:negativetheta3}.}
\end{figure}

\begin{figure}[ht]
\centering
\includegraphics[width=0.7\textwidth]{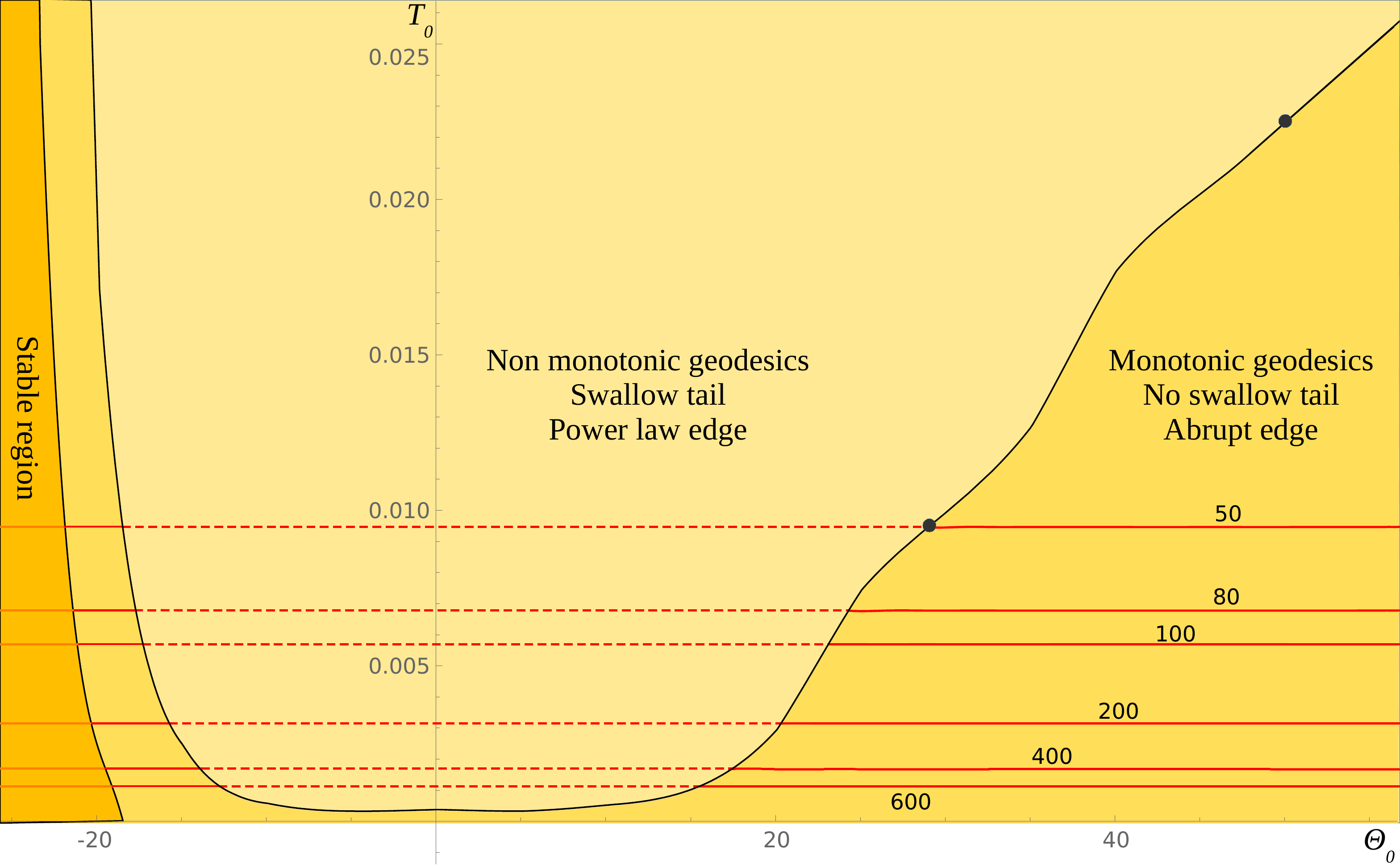}
\caption{ 
Central temperature $\tilde{T}_0$ {\em vs.} central degeneracy $\Theta_0$ phase diagram. 
The horizontal lines correspond to the constant values of $\tilde \mu/\tilde{T}$ for which the free entropy curves in Figs.\ref{fig:massvstemp06} and \ref{fig:massvstemp062} were plotted. The orange, continuous red and dotted red lines are in correlation with Figs. \ref{fig:massvstemp06} and \ref{fig:massvstemp062}.
Solutions are stable only inside the orange region on the left. As $\Theta_0$ is increased, one of the eigenvalues changes its sign and we enter into the unstable regions, depicted in yellow. Further to the right, a second eigenvalue instabilizes in the light yellow area, as the solutions develops a power law edge. In that region, the angular span becomes non-monotonic, and consequently the scalar correlator takes a swallow tail form. For larger values of $\Theta_0$ and well inside the unstable region, the mass as a function of the central density develops a maximum, at the positions depicted by the grey dots for the particular cases of Figs. \ref{fig:positivetheta1} and \ref{fig:negativetheta3}. Coincidentally, the power law edge disappears, together with the swallow tail correlator. Notice that, in particular, this diagram implies that all the plots of Figs. \ref{fig:positivetheta1} and \ref{fig:negativetheta3} correspond to unstable solutions.
}
\label{fig:transitionregion}
\end{figure}

In Figs. \ref{fig:massvstemp06} and \ref{fig:massvstemp062} we plot the thermodynamic curves corresponding to the grand canonical free entropy $\Phi=-\Omega/{\tilde T}$ and minus the mass $-\tilde{M}$, as functions of $1/\tilde{T}$ at fixed $\tilde{\mu}/\tilde{T}$. The central degeneracy $\Theta_0$ plays the role of a parameter. Using the Katz stability criterion, we can differentiate between thermodynamically stable and thermodynamically unstable branches. In Fig. \ref{fig:correlator} we show the angular span $\Delta\varphi$ as a function of the geodesic tip $r_*$ and the on-shell particle action, whose negative exponential gives the scalar correlator, as a function of $\Delta\varphi$.
It is possible to link all the different features detected in the unstable regions involved in Figs. \ref{fig:positivetheta1} to \ref{fig:correlator} by reading from left to right the phase diagram presented in Fig. \ref{fig:transitionregion}.

\renewcommand{\labelenumi}{(\roman{enumi})}
\begin{enumerate}
\item  The transition from thermodynamically stable to unstable solutions, manifested by the first vertical asymptota in Fig. \ref{fig:massvstemp06} and \ref{fig:massvstemp062}, always occurs within the diluted regime, {\em i.e.} for $\Theta_0\sim -20$, for all the constant values of $\tilde \mu/\tilde T$ here analyzed. In this regime, density profiles have an abrupt edge, the angular span $\Delta \varphi$ is monotonic as a function of the tip position $r_*$, and the correlator is well behaved. 

The onset of the instability is denoted in Fig. \ref{fig:transitionregion} by the separatrix between the orange and the yellow regions. There, the central temperature $\tilde T_0$ is well below the critical value $\tilde T_0^{cr}$ at which the turning point in total mass as a function of the central density shows up. Moreover, there is no associated change in the features of the density profiles and the correlators at the transition.
\item For somewhat larger but still negative $\Theta_0$ along any constant $\tilde \mu/\tilde T$ curve in Fig. \ref{fig:transitionregion}, the unstable solutions enter into the light yellow region, characterized by
{\renewcommand{\labelenumi}{(\roman{enumi})}
\begin{enumerate}
\item Radial density profiles with a power law edge, see Fig. \ref{fig:negativetheta3}.
\item A non-monotonic behavior in the angular span $\Delta \varphi$ as a function of the geodesic tip $r_*$, see Figs. \ref{fig:negativetheta3} and \ref{fig:correlator}. 
\item A swallow tail behavior in the scalar correlator as a function of the angular span $\Delta \varphi$, see Fig.  \ref{fig:correlator}. 
\end{enumerate}}
These features persist up to positive values of $\Theta_0$, see Fig. \ref{fig:positivetheta1}. Interestingly, this second transition occurs for $\tilde T_0\ll\tilde T_0^{cr}$, and therefore always prior to the turning point in the total mass as a function of central density. 

\item There always exists a $\Theta_0>0$ along any constant $\tilde \mu/\tilde T$ curve in Fig. \ref{fig:transitionregion}, such that the unstable solutions leave the light yellow region to re-enter into the yellow one. Remarkably, this third transition occurs precisely at the critical temperature $\tilde T_0^{cr}$ at which the maximum in the total mass versus the central density shows up, see Fig. \ref{fig:positivetheta1}. The radial density profiles recover the abrupt edge, the angular span $\Delta\varphi$ as a function of the geodesic tip $r_*$ becomes monotonic again, and the swallow tail behavior in the correlators vanishes.

Such a critical temperature does not occur at any characteristic point in the thermodynamic curves of Figs. \ref{fig:massvstemp06} and \ref{fig:massvstemp062},  though certainly after the second vertical asymptota. 

\item For larger values of $\Theta_0$, the thermodynamic curves of Figs. \ref{fig:massvstemp06} and \ref{fig:massvstemp062} manifest a second spiral behavior, similarly to what was observed in the flat space case in \cite{2018arXiv180801007A}. This is only visible in our plots for the cases $\tilde \mu/\tilde T=50$ and $\tilde \mu/\tilde T=80$ due to the limitations in the parameter space coverage reached in this work.
\end{enumerate}

\newpage
 
\newpage
\section{Discussion}
\label{sec:discussion}
\vspace{-.45cm}
We studied the thermodynamic stability of a holographic neutron star at finite temperature.

As previously reported \cite{2018JHEP...05..118A}, density profiles manifest the rich core - halo structure known from the asymptotically flat case \cite{2019PDU....24..278A}. A dense core and diluted halo are present at large central degeneracies $\Theta_0$, and disappear as the central degeneracy becomes negative. As the central temperature $\tilde T_0$ is increased, the initially sharp end of the star density profile as a function of the radius degenerates into a power law.  This precedes a maximum on the total mass $\tilde{M}$ as a function of the central density, {\em i.e.} a turning point.
We probed the background with the scattering of a massive Euclidean particle, obtaining the spanned angle $\Delta\varphi$ as a function of the minimum radius $r_*$ reached by the particle trajectory. We observed that, at the region of parameters where the density profile has a power law edge, this function is non-monotonic.

From the holographic perspective, we calculated the grand canonical free entropy and perform a Katz equilibrium analysis. We found that the system has a stable branch that destabilizes at a certain temperature and never stabilizes again.  We calculated the holographic two point correlator of a scalar operator in the large conformal dimension limit, and verify that the non-monotonicity of the scattering problem gives rise to a correlator with a non-vanishing angular derivative at an angular separation of $\pi$. Interestingly, such a behavior manifests only inside the unstable branches, where  we also observed bulk profiles with a power law edge, and a turning point in the total mass.

As a main conclusion of the present work, the presence of a power law edge in the density profile, or of a non-monotonic scattering problem, can be taken as a signature of an unstable background from the bulk perspective. In this way, the appearance of a power law characteristic of a continuous phase transition is manifest in the bulk. Conversely, from the holographic point of view, a correlator with a non-vanishing derivative at an angular separation of $\pi$, is an indicator of a phase transition. Since after the onset of the instability the system goes into the gravitational collapse, we identify such phase transition as the confinement to deconfinement one, generalizing former results \cite{DeBoer2009, Arsiwalla2010} to the to finite temperature case. Interestingly, the instability shows up prior to the turning point, a behaviour that is consistent with what is known from flat space \cite{Schiffrin:2013zta, Takami:2011zc, Arguelles:inprep}. 
 
Further steps on this research would be: (i) to investigate the physical reason for  which the swallow tail structure of the correlator shows up coincidentally with the power law edge on the density profiles, (ii) to understand why these behaviors appear inside the unstable branches of the free entropy and prior to the turning point on the mass as a function of the central density; (iii) to reproduce the swallow tail structure out of the large conformal dimension limit, and (iv) to understand the relation between thermodynamical and dynamical instabilities \cite{greco1, greco3}.

\acknowledgments
The authors thank G.A. Silva, D.H. Correa and P.A.G. Pisani for useful comments. This work has been funded by the CONICET grants PIP-2017-1109 and PUE 084 and UNLP grant PID-X791.

\appendix

\section{Details on the bulk state}
\label{sec:background}
\paragraph{The model}
\label{sec:model}

We want to describe the thermodynamics of a very large number of neutral self-gravitating fermions in equilibrium in $3+1$ dimensions in global AdS spacetime. We approximate the dynamics with a perfect fluid coupled to the gravitational field. We use the action
\begin{equation}
S= S_{\sf Gravity}+S_{\sf Fluid}\,.
\label{eq:action}
\end{equation}
Here the gravity part reads
\begin{equation}
S_{\sf Gravity}= \frac{1}{16\pi G}\int d^4x\sqrt{-g}\left(R-2\Lambda\right)\,,
\label{eq:action.gravity}
\end{equation}
where $\Lambda=-3/L^2$ being $L$ the AdS length, and we chose natural units such that $\hbar=c=k_{B}=1$ throughout this work. On the other hand the perfect fluid part can be written as a Schutz action
\begin{equation}
S_{\sf Fluid}= \int d^4x\sqrt{-g}\left(-\rho +\sigma\, u^\mu\left(\partial_\mu\phi+\theta\,\partial_\mu s\right)+\lambda \left(u_\mu u^\mu+1\right)\right)\,,
\label{eq:action.fluid}
\end{equation}
where $\rho$ is the fluid energy density taken as a function of the fluid particle density $\sigma$ and $s$ the entropy per particle, while $u_\mu$ is its four-velocity. The variables  $\phi$, $\theta$ and $\lambda$ are auxiliary Lagrange multipliers, enforcing that the particle number $\sigma$ and the entropy density $\sigma  s$ are conserved, and that the four-velocity $u_\mu$ is a time-like unit vector.
The resulting equations of motion read
\begin{eqnarray}
\sigma\,\left(\partial_\mu\phi+\theta\,\partial_\mu s\right)&=&-\lambda u_\mu\,,
\label{eq:delta.umu}
\\
u^\mu\left(\partial_\mu\phi+\theta\,\partial_\mu s\right)&=&\mu\,,
\label{eq:delta.sigma}
\\
u_\mu u^\mu&=-1\,,
&
\label{eq:delta.lambda}
\\
\partial_\mu\left(\sigma u^\mu\right)&=&0\,,
\label{eq:delta.phi}
\\
\partial_\mu\left(\sigma s\, u^\mu\right)&=&0\,,
\label{eq:delta.theta}
\\
u^\mu\partial_\mu \theta\, &=&-T\,,
\label{eq:delta.s}
\\
G_{\mu \nu}+\Lambda g_{\mu \nu}&=&8\pi G \left(P g_{\mu \nu}+(P+\rho)\,u_{\mu}u_{\nu}\right)\,.
\label{eq:einstein}
\end{eqnarray}
In these equations we have defined the local chemical potential $\mu \equiv {d\rho}/{d\sigma}$, local temperature $\sigma \,T \equiv {d\rho}/{ds}$, and pressure $P \equiv-\rho +\mu\sigma$. The equations in the first two lines fix the auxiliary variables, implying $\lambda=\mu\sigma$ and $\partial_\mu\phi+\theta\,\partial_\mu s=-\mu \,u_\mu$. The third imposes that the four-velocity is unitary and time-like. The next two lines correspond to the conservation of the particle current $\sigma u^\mu$ and the entropy current $\sigma s u^\mu$. Then we define the temperature as the proper time derivative of the ``thermasy'' $\theta$. The last line contains the Einstein equations.

The above equations of motion need to be supplemented with a explicit dependence of $\rho$ in $\sigma$ and $s$. In the limit $mL \gg 1$, in which there is a large number of particles within one AdS radius \cite{DeBoer2009}, we can define such relation implicitly by
\begin{eqnarray}
\label{eq:density}
\rho&=&\frac{g}{8\pi^3}\int f(p)\,\sqrt[•]{p^2+m^2}\,d^3p,
\\
\label{eq:pressure}
P&=&\frac{g}{24\pi^3}\int f(p)\,\frac{p^2}{\sqrt[•]{p^2+m^2}}\,d^3p,
\end{eqnarray}
where $g$ is the number of fermionic species or spin degeneracy, the integration runs over all momentum space, and $f(p)$ is the Fermi distribution for a fermion of mass $m$ with local temperature  $T$ and local chemical potential $\mu$
\begin{equation}
f(p)=\frac{1}{e^{\frac{\sqrt[•]{p^2+m^2}-\mu}{T}}+1}\,.
\end{equation}
\paragraph{The Ansatz}
\label{sec:ansatz}
We solved the above equations of motion with a stationary spherically symmetric ``neutron star'' Ansatz with the form
\begin{eqnarray}\label{metric}
&&ds^{2}=L^{2}(-e^{\nu(r)}dt^{2}+e^{\lambda(r)}dr^{2}+r^2d\Omega^2_2),
\\
&&u^\mu = u(r)\partial_t\,,
\end{eqnarray}
where $d\Omega^2_2=d\vartheta^2+\sin^2\!\vartheta\, d\varphi^2$ is a two-sphere. The local temperature $T$ and chemical potential $\mu$ are then radial functions.
Defining the functions $\tilde M$ and $\chi$  by
\begin{equation}
e^{\lambda}=\left( 1-\frac{2\tilde{M}}{r} +r^2 \right)^{-1}\,,
\end{equation}
\begin{equation}
e^{\nu}=e^{\chi } \left( 1-\frac{2\tilde{M}}{r} +r^2 \right)\,,
\end{equation}
equations \eqref{eq:delta.umu}-\eqref{eq:einstein} become the thermodynamic equilibrium conditions of Tolman and Klein
\begin{equation}\label{eq:tolman}
e^{\frac{\nu}{2}}\tilde{T}={\rm constant}\,,
\end{equation}
\begin{equation}\label{eq:klein}
e^{\frac{\nu}{2}}\tilde{\mu}={\rm constant}\,,
\end{equation}
and the Einstein equations
\begin{equation}\label{eq:M}
\frac{d\tilde{M}}{d r}=4\pi r^2\tilde{\rho},
\end{equation}
\begin{equation} \label{eq:nu}
\frac{d {\chi}}{dr}=8\pi r\left(\tilde{P}+\tilde{\rho}\right)e^{\lambda}\,.
\end{equation}
In the above expressions to we have written the dimensionless combinations $\tilde{T}= T/m$ and $\tilde{\mu}=\mu/m$ for the temperature and chemical potential. The constants
are obtained by evaluating the expression
 at a reference point, conventionally taken at $r=0$. Moreover, we have re-scaled the density $\tilde \rho=GL^2 \rho$ and pressure $\tilde P=GL^2 P$. They are obtained from \eqref{eq:density} and \eqref{eq:pressure} rewritten in terms of the variable $\epsilon=\sqrt{1+p^2/m^2}$, resulting in the expressions
\begin{equation}
\label{eq:density.dimensionless}
\tilde \rho=\gamma^2\int_1^\infty
\frac{\epsilon^2\sqrt{\epsilon^2-1}}{e^{\frac{\epsilon-\tilde{\mu}}{\tilde{T}}}+1}\,d\epsilon\,,
\end{equation}
\begin{equation}
\label{eq:pressure.dimensionless}
\tilde P=
\frac{\gamma^2}3\int_1^\infty
\frac{\left(\sqrt{\epsilon^2-1}\right)^3}{e^{\frac{\epsilon-\tilde{\mu}}{\tilde{T}}}+1}\,d\epsilon\,,
\end{equation}
where $\gamma^2=gGL^2m^4/2\pi^2$.

By expanding the equations \eqref{eq:M} and \eqref{eq:nu} around the center of the configuration $r=0$, we obtain the boundary conditions that correspond to a regular metric, as
\begin{eqnarray}
\tilde{M}(0)=0,
\label{eq:boundary.mass}\\
\chi(0)=0,\\
\tilde{T}(0)=\tilde{T_{0}},\\
\tilde{\mu}(0)=\mu_{0}\equiv \Theta_{0}\tilde{T_{0}}+1.
\label{eq:boundary.mu}
\end{eqnarray}
Here $\Theta_{0}$ is called the ``central degeneracy'', and we used it as a way to parameterize the central chemical potential. Families of solutions are then indexed by the parameters $(\tilde T_0, \Theta_0. \gamma^2)$.

\newpage
\section{Details on space-like geodesics}
\label{subsec:geodesics}

In order to probe the resulting gravitational background, we study the scattering of a massive particle with an Euclidean worldline. The trajectories are obtained by extremizing the Euclidean particle action
\begin{equation}
S^E_{\sf Particle}= {\sf m}\int d\tau_E\,\sqrt{g_{\mu\nu} \, {x'}^\mu {x'}^\nu}\,.
\label{eq:action.particle.no.gauge}
\end{equation}
Here ${\sf m}$ is the mass of the particle, $x^\mu(\tau_E)$ describes the geodesic in terms of an Euclidean parameter $\tau_E$, and a prime ($'$) means a derivative with respect to ${\tau_E}$.
Evaluated on our Ansatz  \eqref{metric} this action  reads
\begin{equation}
S^E_{\sf Particle}= {\sf m}L\int d\varphi\,\sqrt{r^2+e^{\lambda(r)}r'^{2}}\,,
\label{eq:action.particle.gauge.fixed}
\end{equation}
where we specialized to constant time $t'=0$ and chose coordinates such that the trajectory lies in the equator of the sphere $\vartheta=\pi/2$. Moreover we fixed the reparametrization gauge as $\tau_E=\varphi$.
The resulting system is invariant under shifts of the Euclidean parameter $\varphi$, resulting in a conserved quantity
\begin{equation}
r_* = \frac{r^2}{\sqrt{r^2+e^{\lambda(r)}r'^{2}}}\,.
\label{eq:r.star}
\end{equation}
The value of $r_*$ coincides with the value of the coordinate $r$ at tip of the trajectory, defined as the point at which $r'=0$. It can be used to label different geodesics by their minimum approach $r_*$ to the center of the geometry.

%
%

We are interested in geodesics starting at a very large radius $r_\epsilon$, falling into the geometry up to a minimum radius $r_*$, and then bouncing back into the asymptotic region, spanning a total angle $\Delta \varphi$ (see Fig.\ref{fig:sphere}). Then, our boundary conditions are
\begin{equation}
\left.r\right|_{\varphi=0}=\left.r\right|_{\varphi=\Delta\varphi}=r_\epsilon\,,
\end{equation}
here $\Delta\varphi$ is can be calculated according to
\begin{equation}
\Delta\varphi = \int d\varphi=\int \frac{dr}{r'}\,.
\end{equation}
Solving \eqref{eq:r.star} for the velocity $r'$ and using the fact that the trajectory is symmetric around its tip, we get $\Delta\varphi$ as a function of $r_*$, in the form
\begin{equation}
\Delta\varphi = 2r_*\int_{r_*}^{r_\epsilon} dr\,\frac{ e^{\frac{\lambda(r)}2}}{
r \sqrt{
r^2-r_*^2
}
}\,.
\label{eq:delta.varphi}
\end{equation}
For $r_*=0$ we have no scattering and then $\Delta\varphi=\pi$. On the other hand in the limit of very large  $r_*$ we get $\Delta\varphi=0$, implying a backward scattering. In the intermediate region, the behaviour of $\Delta\varphi$ can be either monotonic or non-monotonic. In the last case, the same angle $\Delta\varphi$ is spanned by geodesics with different values of the minimum approach radius $r_*$.

\begin{figure}[ht]
\centering
\includegraphics[width=0.7\textwidth]{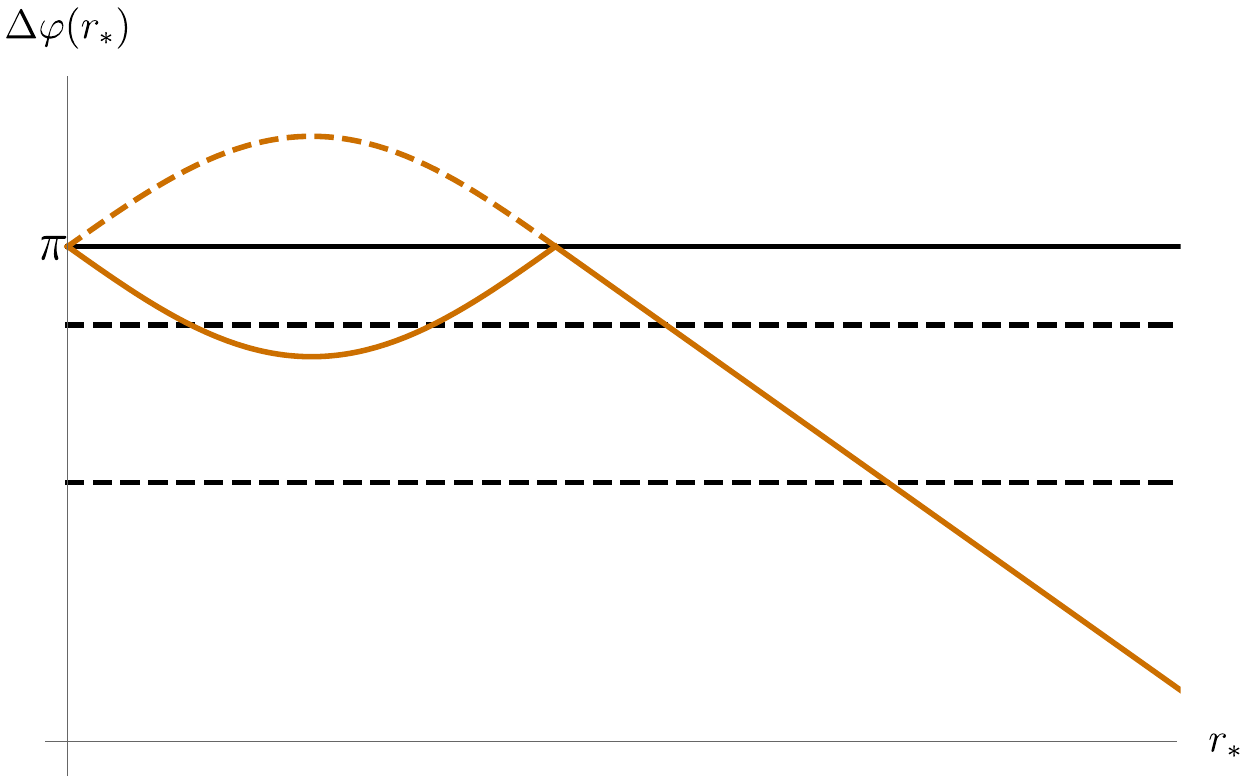}
\put(-130,112){\scriptsize $\Delta \varphi$ with multiple values of $r_*$ \normalsize}
\put(-281,81){\scriptsize $\Delta \varphi$  with a single value of $r_*$ \normalsize}
\caption{\label{fig:angle-generic} The angle $\Delta \varphi$ as a function of the tip position $r_*$. Whenever the integral \eqref{eq:delta.varphi} returns a value larger than $\pi$ (dotted curve), it must be mapped to $2\pi -\Delta\varphi$ (continuous curve). Notice that this results in a non-monotonic behavior, with three different values of $r_*$ returning the same value of $\Delta\varphi$, for $\Delta\varphi$ close enough to $\pi$.
}
\end{figure}

Due to the symmetries of the problem, a value of $\Delta\varphi$ larger than $\pi$ is not physically different from the value smaller than $\pi$ obtained by the transformation $\Delta\varphi\to 2\pi-\Delta\varphi$. This is evident in Fig.\ref{fig:sphere}. Thus, in what follows we restrict $\Delta\varphi$ to values smaller than $\pi$, applying such transformation whenever the integral \eqref{eq:delta.varphi} returns a value larger than $\pi$. This may result in a non-monotonic behaviour of $\Delta\varphi$ as a function of $r_*$, see Fig.\ref{fig:angle-generic}.

\newpage 
\section{Details on the correlator of a scalar operator}
\label{sec:correlators}
In order to probe the field theory on the boundary, we consider  a scalar operator, and we used the dictionary of the AdS/CFT correspondence to evaluate its two-point correlator. We consider two points in the boundary sphere, separated by an angle $\Delta \varphi$ at fixed Euclidean time. In the limit of a large conformal dimension $\Delta\equiv {\sf m}L$, the correlator is given as
\begin{equation}
\langle{\cal O}(\Delta\varphi){\cal O}(0)\rangle=\lim_{r_{\epsilon}\rightarrow\infty}r_{\epsilon}^{2{\sf m}L}
e^{-S^{E~ \sf on-shell}_{\sf Particle}(\Delta\varphi)},
\label{eq:correlator}
\end{equation}
where the exponent is the Euclidean particle action \eqref{eq:action.particle.no.gauge} evaluated on shell, and $r_{\epsilon}$ is an UV bulk regulator, whose power is included in the pre-factor in order to get a finite result. Solving \eqref{eq:r.star} for $r'$ and plugging back into \eqref{eq:action.particle.gauge.fixed}, we obtain the Euclidean action as
\begin{equation}
S^{E~ \sf on-shell}_{\sf Particle}=2{\sf m}L\int_{r_{*}}^{r_{\epsilon}}dr
\frac{r e^{\frac{\lambda(r)}{2}}}{r \sqrt{r^2-r_{*}^2}},
\label{eq:action.particle.on.shell}
\end{equation}
where we have included the same cutoff $r_{\epsilon}$.

Equation \eqref{eq:action.particle.on.shell} gives the on-shell action as a function of the position of tip if the geodesic $r_*$. Equation \eqref{eq:delta.varphi} on the other hand, provide $\Delta\varphi$ as a function of the same variable. This allows us to parametrically plot the the on-shell action as a function of the angular separation. Interestingly, for the cases in which $\Delta\varphi$ is non-monotonic as a function of $r_*$ that we studied in section \ref{subsec:geodesics}, there are three values of $S^{E~ \sf on-shell}_{\sf Particle}$ for a single value of $\Delta \varphi$, {\em i.e.} the Euclidean particle action become multi-valued. The correlator is then given by the absolute minimum of the Euclidean particle action, which corresponds to the smaller branch of the multi-valued function.

\end{document}